\documentclass{aa}  

\usepackage{graphicx}
\usepackage{newtxtext,newtxmath}
\usepackage{subcaption}         
\usepackage{lscape}             
\usepackage{bm}
\usepackage{amsmath}
\usepackage{amssymb}
\usepackage{natbib}

\usepackage{xcolor}
\usepackage{placeins}           

\usepackage{hyperref}
\bibpunct{(}{)}{;}{a}{}{,}

\begin{document}

    \title{Impact of embedded circumplanetary winds on the circumstellar disk}

    \subtitle{I. Reshaping the local accretion environment}

    \author{
        Danilo~Sep\'ulveda-Rojas\inst{\ref{UCH}}
        \and
        Pablo~Ben\'itez-Llambay\inst{\ref{UAI}}
        \and
        Simon~Casassus\inst{\ref{UCH},\ref{DO}}
        }

    \institute{Departamento de Astronom\'{\i}a, Universidad de Chile, Casilla 36-D, Santiago, Chile\\
    \email{danilo.sepulveda@ug.uchile.cl} \label{UCH}
    \and
    {Facultad de Ingenier\'ia y Ciencias, Universidad Adolfo Ib\'a\~nez, Av. Diagonal las Torres 2640, Pe\~nalol\'en, Chile} \label{UAI}
    \and
    {Data Observatory Foundation, Eliodoro Y\'a\~{n}ez 2990, Providencia, Santiago, Chile} \label{DO}
    }

    \date{Received 29 October 2025; Accepted 6 February 2026}

    
    \abstract
    {The existence of winds is among the uncertainties related to the growth of giant planets. Such circumplanetary outflows have been proposed to explain kinematic and chemical structures in protoplanetary disks. }
    {We investigate the immediate impact of circumplanetary outflows on the circumstellar disk environment, the planetary vicinity, and planetary growth.}
    {We performed three-dimensional hydrodynamic simulations using \texttt{FARGO3D}, implementing a parametric wind launched from the vicinity of an embedded planet.}
    {Although the imposed configurations for the outflows do not significantly alter the global structure of the disk, they do substantially redistribute material in the vicinity of the embedded planet. In particular, the wind redirects accretion flows from polar to equatorial latitudes, resulting in variable accretion patterns over time. Although the mass accretion rate variations depend on the efficiency of the outflows, their presence diminishes the accretion rate over time and the total mass reservoir within the Hill sphere and the planet's direct vicinity, potentially slowing or limiting planetary growth.}
    {}


    \maketitle

\nolinenumbers
\section{Introduction} \label{sec:introduction}
In astrophysics, accretion processes onto compact bodies are systematically associated with jets and outflows, a phenomenon that has been well-documented over a wide range of mass scales \citep[e.g.,][]{accretion_frank2002}, from young stars ($\sim M_{\odot}$) to supermassive black holes ($\sim 10^{6} M_{\odot}$). However, only a handful of articles have addressed the possibility of planetary outflows. Early work by \citet{Fendt2003} provided an analytical prescription for the physical conditions required to launch an MHD-driven planetary outflow. They proposed that magnetic forces can lift a fraction of the accreting magnetized plasma to higher altitudes, where it would couple with the large-scale magnetic field, launching a slow wind. The material would then be further accelerated by magneto-centrifugal forces and, at greater distances, by the Lorentz force. A key parameter in this mechanism is the disk magnetization, which governs the efficiency and structure of the outflow. Building on this framework, \citet{Machida2006} used ideal MHD simulations in a shearing box to explore the impact of different disk magnetization levels, showing that even weak magnetic fields can lead to outflows in the azimuthal direction of the disk. Their simulations revealed well-collimated, "tubelike" outflows forming along the azimuthal direction of the disk, accompanied by equatorial accretion flows. Both studies predicted and observed field strengths of up to 10 gauss, with outflow velocities ranging from $10~\mathrm{km\,s^{-1}}$ to $\sim 60~\mathrm{km\,s^{-1}}$.

Further supporting these findings, \citet{Gressel2013GLOBAL} reported the launching of relatively diffuse, jet-accelerated winds originating from the conical regions above and below the circumplanetary disk (CPD) in global, non-ideal MHD simulations. Similarly, in the non-ideal MHD simulations by \citet{Wafflard-Fernandez2023_planetDiskWindInteraction}, a sporadic and very collimated jet develops for the case of thermal-mass bodies. Broadly speaking, these outflows appear to be a common feature in MHD environments, characterized by strong, collimated winds launched within the inner half of the Hill sphere ($r < r_{\mathrm{H}}/2$) and extending along the meridional direction in the planet's frame of reference.

Radiative feedback from accretion heating has also been identified as a mechanism capable of driving planetary outflows. \citet{Chrenko2019} showed that the heat generated by pebble accretion can produce sufficient thermal pressure to launch outflows from the planet. By incorporating accretion luminosity into their simulations and departing from the commonly assumed isothermal conditions, they identified the emergence of cyclic, unsteady outflows localized around the planet and extending over a few Hill radii. The resulting flow geometry was irregular and largely uncollimated, governed by local thermal instabilities such as those described by the Schwarzschild criterion. A similar behavior was later found by \citet{Chrenko2023_AccretingLum} using a constant opacity, where variations were again accompanied by vertical outflows, and once more in \citet{Chrenko2025_CObubble} involving multiple species. Another alternative mechanism was found by \cite{Bailey2021_FlowFieldEccentricOrbits}, who demonstrated using an isothermal 3D model that vertical outflows should be common for mildly eccentric planets, although at very localized scales (about one Hill radius).

The appearance of planetary outflow-like structures under different physical conditions suggests that such a process could be relatively common in young systems and may arise in a variety of ways, each with its own implications and timescales. Since these processes are mainly driven by accretion, it underscores the importance of exploring whether planetary outflows could impact the planet formation and growth process. Including this momentum source in the system could alter the dynamics and structures of the material surrounding the planetary region, potentially altering the accretion rates and the accretion direction. On top of this, outflows could imprint structural variations in the parent disk and its dynamics. However, a key difficulty for this goal lies in the lack of knowledge and constraints on the physical conditions required to drive these outflows. Their strong dependence on the local environment adds to the complexity of the driving mechanisms, resulting in a wide and degenerate parameter space. Their fundamental properties, such as shape, strength, and variability, are poorly understood, as is how they correlate with the physical conditions of the system, which complicates any self-consistent study.

Motivated by the theoretical predictions, numerical results, and potential implications described above, we investigate the hypothesis that embedded planets may drive localized, medium-scale outflows. In this work, rather than committing to a specific launching mechanism, we employ 3D ideal-hydrodynamic simulations using a parametric model for the outflow, introduced manually into a non-magnetized setup and detailed in Sect.~\ref{sec:numerical}. This approach allows us to isolate and examine the effects of an active planetary wind on the surrounding disk by directly controlling key properties such as its geometry, strength, and spatial extent, without the complexity of their dependence on environmental conditions. Within this framework, we focus on the direct effects of its presence on the disk structure and in the planet's vicinity, and its implications for the inflow of material. Although we do not directly address the detectability of these outflows, their presence could introduce perturbations in the disk that, among other things, could lead to Doppler-flip signatures. The ultimate goal is to evaluate the role planetary outflows could play in planetary growth and in shaping disk evolution.

\section{Numerical methods} \label{sec:numerical}
This section provides an overview of the numerical framework used to model planetary outflows and accretion processes. We describe the essential equations that govern this physical problem and outline the numerical methods and prescriptions that define our approach for implementing these phenomena in hydrodynamical simulations.

\subsection{Simulation setup}
We model a 3D gas disk (with no dust species) using a spherical coordinate system $(\varphi, r, \theta)$ centered on a star of mass $M_{\star}$. The code is based on the \href{https://github.com/FARGO3D/fargo3d/tree/860dcf1828a09d16ed3fb267c456c5cfc0aa76ef}{2024 public version} of \texttt{FARGO3D} \citep{Benitez2016FARGO3D}, with modifications to include our wind prescription.

\subsubsection{Basic equations}
For the purposes of our simulations, the hydrodynamical model is employed to solve the conservation of mass (Equation~\eqref{eq:continuity}) and conservation of momentum (Equation~\eqref{eq:momentumConservationBasic}) for the gas only,

\begin{align}
    \partial_t \rho_g + \nabla \cdot (\rho_g \mathbf{v}_g) &= 0,
    \label{eq:continuity} \\
    \partial_t \mathbf{v}_g + \mathbf{v}_g \cdot \nabla \mathbf{v}_g
    &= -\frac{\nabla P}{\rho_g} - \nabla \Psi + \frac{\nabla \cdot \boldsymbol{T}}{\rho_g},
    \label{eq:momentumConservationBasic}
\end{align}

where $\rho_g$ is the gas volumetric density, $\mathbf{v}_g$ is the velocity vector of the gas, $P$ is the gas pressure, $\Psi$ represents external potentials, and $\boldsymbol{T}$ is the viscous stress tensor, given by:
\begin{equation}
    \boldsymbol{T} = \rho_g \nu \left[ \nabla \mathbf{v}_g + (\nabla \mathbf{v}_g)^T - \frac{2}{3}(\nabla \cdot \mathbf{v}_g) \mathbf{I} \right],
    \label{eq:viscous_tensor}
\end{equation}
with $\nu$ the kinematic viscosity and $\mathbf{I}$ the identity matrix. To model the viscosity in the disk, we employ the $\alpha$-viscosity model described in \cite{1973A&A....24..337S}, with a value of $\alpha = 10^{-4}$. The kinematic viscosity is given by:
\begin{equation}
    \nu = \alpha c_s H,
    \label{eq:viscosity}
\end{equation}
where $c_s$ is the sound speed and $H$ is the pressure scale height of the disk, which we define as a constant fraction of the radius, $H(r) = h_0 (r/R_0)$, with an aspect ratio of $h_0=0.05$.

The wind prescription, as described later in Sect.~\ref{sec:wind_prescription}, is an imposed acceleration term, which we refer to as $\mathbf{\Gamma}$, and thus must be included in Equation~\eqref{eq:momentumConservationBasic}. Similarly, the gravitational accelerations from the planet, $\mathbf{G_p}$, and the star, $\mathbf{G}_{\star}$, must be considered. This leads to Equation~\eqref{eq:momentumConservationWind} as the momentum equation in our setup. We have separated the external potential gradient $-\nabla \Psi$ from Equation~\eqref{eq:momentumConservationBasic} into the stellar gravity $\mathbf{G}_{\star}$ and planetary gravity $\mathbf{G_p}$, and added the wind acceleration $\mathbf{\Gamma}$.
\begin{equation} \label{eq:momentumConservationWind}
    \partial_t \mathbf{v}_g + \mathbf{v}_g \cdot \nabla \mathbf{v}_g = -\frac{\nabla P}{\rho_g} + \mathbf{G}_{\star}~(r) + \mathbf{G_p}~(r^{\prime}) + \mathbf{\Gamma}~(r^{\prime}) + \frac{\nabla \cdot \boldsymbol{T}}{\rho_g}
\end{equation}
where the radial distance in spherical coordinates, $r$, is measured relative to the central star, while $r^{\prime}$ is measured relative to the planet. The pressure is given by $P=c_s^2 \rho_g$, which corresponds to a locally isothermal disk. In Equation~\eqref{eq:momentumConservationWind}, $\mathbf{\Gamma}$ (an outward acceleration) and $\mathbf{G_p}$ (an inward acceleration) naturally compete, as reflected in the work integral in Equation~\eqref{eq:energy_balance}.
The planet follows a fixed circular, planar orbit around the central star at a radial distance $r_{\rm p} = R_0$. Its gravitational field is modeled using a Plummer potential with a smoothing length of approximately 3 grid cells. To ensure the correct treatment of the non-inertial reference frame, we explicitly include the indirect acceleration term corresponding to the force exerted by the planet on the central star. The planetary potential is introduced instantaneously at $t=0$ without a temporal ramp-up phase. While this abrupt introduction may generate brief initial transients in the accretion flow, the simulations extend for 500 orbits, providing enough time for the system to relax into a quasi-steady state. Consequently, any artifacts associated with the initialization are short-lived and do not significantly impact the long-term accretion efficiency or global disk dynamics.

Our disk model corresponds to one in which the disk is vertically isothermal on curves of constant spherical radius\footnote{This model is implemented in the FARGO3D setup \texttt{p3diso}.}, for which the steady-state solution implies a temperature profile and volumetric density prescribed by Equations \eqref{eq:setup_tempProfile} and \eqref{eq:setup_densityProfile}, respectively \citep{masset2016}. The model assumes a surface density $\Sigma_0$ at the reference radius $R_0$, which is set to the planet's semi-major axis.

\begin{align}
    T(r)
    &= \frac{2.3 m_H}{k_B} \frac{h_0^2 G M_{\star}}{r}
    \approx 61.8 \left(\frac{10 AU}{r}\right)
    \left(\frac{M_{\star}}{M_{\odot}}\right)\, \mathrm{K},
    \label{eq:setup_tempProfile} \\
    \rho(r, \theta)
    &= \frac{\Sigma_0}{\sqrt{2 \pi} R_0 h_0}
    \left(\frac{r}{R_0}\right)^{-1.5}
    \sin(\theta)^{-2.5 + 1/h_0^2},
    \label{eq:setup_densityProfile}
\end{align}

where $h_0$ is the aspect ratio at the reference radius $R_0$, $m_H$ is the atomic mass of hydrogen, $k_B$ is the Boltzmann constant, and $G$ is the gravitational constant.

Note that Equation~\eqref{eq:setup_tempProfile} describes the viscous temperature of the disk, not the radiation temperature, as our setup does not implement stellar irradiation in order to maintain scalable results. This choice is physically justified by adopting a reference radius of $R_0 = 10\,\mathrm{AU}$, a region where viscous heating is expected to dominate over stellar irradiation. At this distance, the resulting disk temperatures are consistent with viscously heated models, ensuring that the neglect of irradiation does not introduce a significant physical inconsistency. The adopted reference radius therefore places the simulations in a regime where the simplified physics assumed in this work provide a reasonable approximation of reality and allow for a more robust interpretation of the resulting wind dynamics.

\subsubsection{Numerical grid and boundary conditions}

We limit the wind prescription to a radius of $r_{\mathrm{H}}/2$ to confine its direct effects to the planet's local environment, with $r_{\mathrm{H}}$ the Hill radius, given by

\begin{equation}\label{eq:hill_radius}
    r_{\mathrm{H}} = R_0\left( \frac{M_p}{3M_{\star}}\right)^{1/3},
\end{equation}
where $M_p$ is the mass of the planet.

The wind's momentum is transmitted to the surrounding disk via ram pressure. To spatially resolve the wind on various scales, a high grid resolution is required around the planet and throughout the disk. With this in mind, we chose the grid resolution presented in Table~\ref{tab:grid_parameters}.

\begin{table}[!h]
    \caption{Simulation grid parameters in spherical coordinates.}
    \label{tab:grid_parameters}
    \centering
    \begin{tabular}{ccccc}
        \hline
        \hline
        \textbf{Coordinate} & \textbf{Resolution} & \textbf{Min} & \textbf{Max} & \textbf{Spacing} \\
        \hline
        $\varphi$ & 768 & $-\pi$ & $\pi$ & linear \\
        $r$ & 364 & 0.4 & 2.5 & logarithmic \\
        $\theta$ & 144 & $\pi/2-\eta$ & $\pi/2+\eta$ & linear \\
        \hline
    \end{tabular}
\end{table}

The parameter $\eta$ is defined as $\eta = \arctan(10h_0)$, which ensures the grid resolves the disk vertically up to 10 scale heights from the midplane at the planet's orbital radius. This configuration yields a resolution of approximately 11 cells across the Hill radius in the colatitudinal direction and provides about 5 cells to resolve the wind's cutoff radius ($r_{\mathrm{H}}/2$) in the radial direction.

We apply periodic boundary conditions in the azimuthal direction ($\varphi$). For the radial boundaries, we use a non-reflecting condition to prevent wave reflection from the edges of the grid. In the polar direction ($\theta$), we use an outflow condition at the upper and lower boundaries, allowing material to leave the computational domain freely.

\subsection{Wind prescription} \label{sec:wind_prescription}

As previously introduced, there are multiple physical mechanisms that can drive wind-like ejections from an embedded planet, such as MHD or accretion heating. Yet, despite their physical plausibility, there are few constraints on how the physical parameters of the system would correlate with the generated outflow, complicating a general analytical study of the effects of such phenomena. For this reason, we adopt a complementary approach: we implement a parameterized prescription for a circumplanetary disk wind (CPD wind), motivated by the accretion-driven outflows commonly observed across different astrophysical contexts. Rather than deriving the wind from a specific launching mechanism, we represent it as an imposed acceleration field acting on the gas, analogous to what might arise from ram pressure forces in an expanding shell. Following the suggestion of \citet{Casassus2022Doppler}, we define this expansion pulse as a volumetric force per unit mass, $\mathbf{\Gamma}$, associated with a ram pressure term $\rho v_s^2$ inside a shell of density $\rho$, velocity $v_s$, width $\Delta r_s$, and radial distance $r_s$ from the planet, with $\Delta r_s \sim r_s$. This yields a compact form for the imposed acceleration:

\begin{equation}
    \Gamma = \frac{v_s^2}{r_s}.
    \label{eq:ramPressure_basic}
\end{equation}

\begin{figure}[!ht]
    \centering
    \includegraphics[width=\columnwidth]{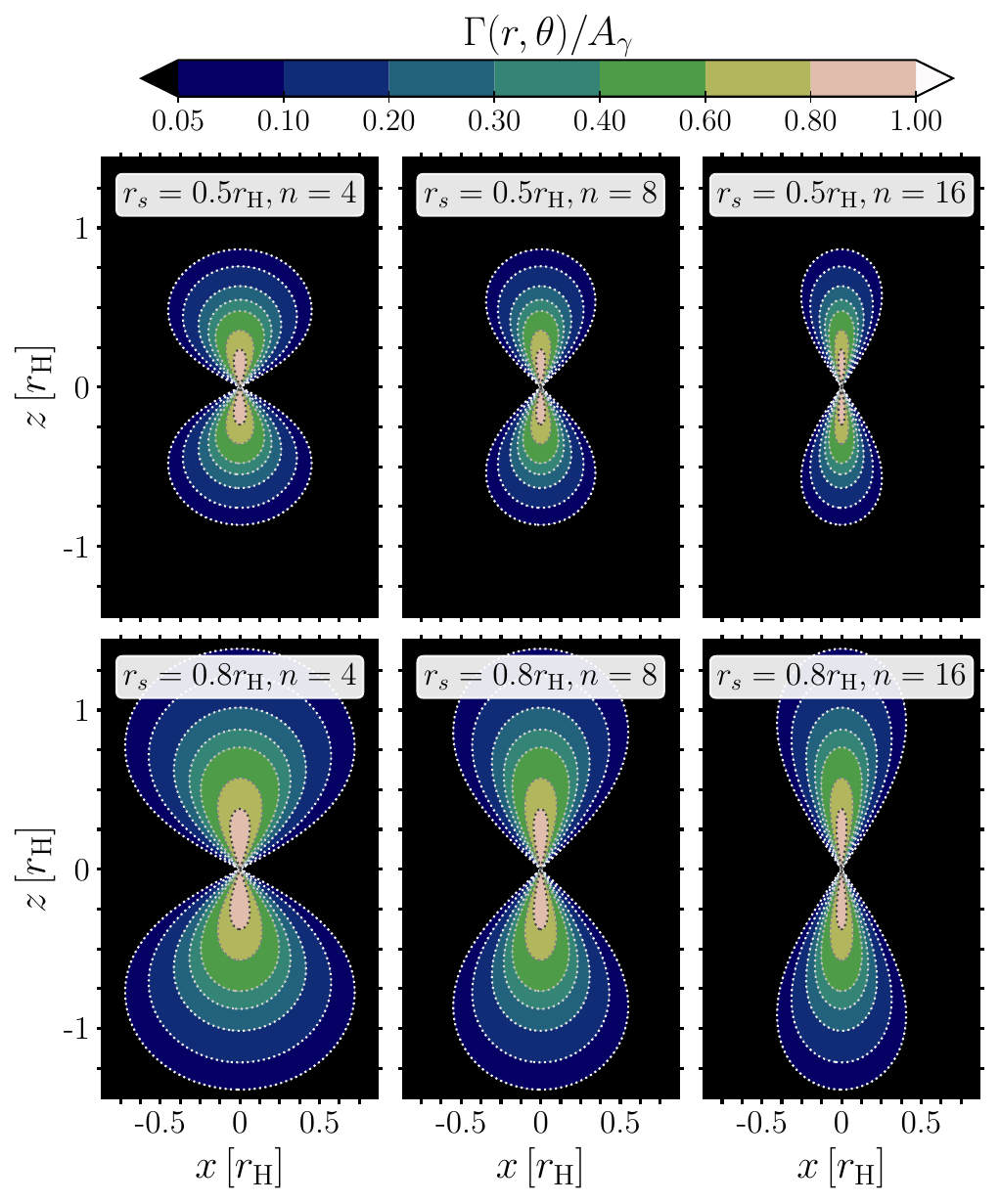}
    \caption{Visual representation of Equation~\eqref{eq:GammaDef}, showing the ratio $\vec{\Gamma}(r, \theta)/A_{\gamma}$ for two smoothing lengths, $r_s = 0.5r_H$ (top row) and $r_s = 0.8r_H$ (bottom row), and three steepness parameters, $n = 4$, $8$, and $16$ (left to right). Colored contours correspond to values of $0.8$, $0.6$, $0.4$, $0.3$, $0.2$, $0.1$, and $0.05$. The Hill radius $r_H$ sets the radial scale, with the planet positioned at the center of each panel.}
    \label{fig:GraphicOfGamma}
\end{figure}

The outflow or wind in this work is formulated as follows:

\begin{equation}
    \mathbf{\Gamma}(r', \theta'') = \frac{GM_{\star}}{R_0^2} A_{\gamma} e^{-r'^2/r_s^2} \cos^n{\theta''} \  \hat{r}',
    \label{eq:GammaDef}
\end{equation}

where $\hat{r}^{\prime}$ is the radial vector relative to the planet's position. The parameters $n$ and $r_s$ control the shape (collimation) and spatial extent of the wind, respectively, as illustrated in Figure~\ref{fig:GraphicOfGamma}. The coefficient $A_{\gamma}$ sets the amplitude (strength) of the outflow, thereby influencing the associated mass-loss rate. Note that $\mathbf{\Gamma}$ is expressed in units of the planet's orbital acceleration $\left(GM_{\star}/R_0^2\right)$, which is the natural unit of acceleration in our hydrodynamical simulations. 

Equation~\eqref{eq:GammaDef} represents a bipolar wind that resembles typical astrophysical outflows. While the exponential term ensures a rapid spatial decay, the function technically extends to infinity, which contradicts the assumption of a localized wind source. We therefore impose a cutoff radius at $r'=r_{\mathrm{H}}/2$, beyond which the outflow acceleration is set to zero (i.e., $\mathbf{\Gamma} = 0$ for $r'>r_{\mathrm{H}}/2$). This confines the wind to the planet's immediate vicinity, from where it transmits its momentum to the rest of the disk through ram pressure.

It can be proven that Equation~\eqref{eq:GammaDef} reaches its maximum effect against the planet's gravity at $r'=r_s$, making the optimal choice for $r_s$ coincide with the cutoff radius, resulting in $r_s=r_H/2$. The collimation parameter $n$ is set to 8, as higher values would require significantly greater resolution to accurately capture the resulting narrow structure, which would exceed the limits of our available computational resources.

Equation~\eqref{eq:GammaDef} is defined in a reference frame centered on the planet, whereas the simulations are carried out in a frame centered on the star. Therefore, to incorporate the wind potential into our simulations, we must introduce the transformation between these two coordinate systems. The details of this transformation are provided in Appendix~\ref{sec:wind_parameterization}.

\subsection{Physical interpretation of $A_{\gamma}$}\label{section:physical_interpretation_of_Ag}

The analysis in Appendix~\ref{section:Strength parameter} establishes the conditions under which a wind can overcome planetary gravity and eject material. While this defines the dynamical regimes of the outflow, it is also useful to connect the wind strength parameter, $A_{\gamma}$, to the magnitude of the kinematic perturbation it induces in the disk. To do this, we perform a simplified analytical estimate.

Consider a parcel of gas accelerated by the wind. The velocity, $v_w$, it acquires can be interpreted as a kinematic perturbation. We can express this perturbation as a fraction, $f$, of the local Keplerian velocity, $v_k$, such that $v_w = f v_k$. The kinetic energy gained by the parcel results from the net work done by the wind acceleration, $\Gamma$, and the planet's gravitational term, $G_p$.

Following the logic of Eq.~\eqref{eq:energy_balance}, we consider a parcel starting at an initial radius $r_1$ and reaching a final radius $r'$ (with $r'>r_1$). A wind can only accelerate material outward once the wind acceleration equals or exceeds the planetary gravitational force. We therefore define $r_1$ as the radius at which this balance occurs (Condition 1 in Appendix~\ref{section:Strength parameter}, $\Gamma(r_1)/G_p(r_1) = 1$). The resulting velocity $v_w$ at radius $r'$ is computed by integrating the net acceleration:

\begin{equation}
    \begin{aligned}
 \frac{1}{2}v_w^2(r_1, r') &= \int_{r_1}^{r'} \left[ A_{\gamma} \frac{GM_\star}{R_0^2} e^{-x^{2}/r_s^2} - \frac{GM_p}{x^{2}} \right] dx \\
        &= A_{\gamma} \frac{GM_\star}{R_0^2} \frac{\sqrt{\pi} r_s}{2} \left[ \text{erf}\left(\frac{r^{\prime}}{r_s}\right) - \text{erf}\left(\frac{r_1}{r_s}\right) \right] \\
        &\quad + GM_p\left( \frac{1}{r^{\prime}}-\frac{1}{r_1} \right).
    \end{aligned}
\end{equation}
Normalizing $v_w^2(r_1, r')$ by the squared Keplerian velocity, $v_k^2 = GM_\star/R_0$, where $R_0$ is the radial location of the planet, yields the expression for $f^2$:

\begin{equation}
    \begin{aligned}
 f^2(r') &= A_{\gamma}\frac{\sqrt{\pi}r_s}{R_0} \left[ \text{erf}\left(\frac{r'}{r_s}\right) - \text{erf}\left(\frac{r_1}{r_s}\right) \right] + \frac{2M_p R_0}{M_{\star}} \left( \frac{1}{r'}-\frac{1}{r_1} \right).
    \label{eq:f_from_Agamma}
    \end{aligned}
\end{equation}

\begin{figure}[!ht]
    \centering
    \includegraphics[width=\columnwidth]{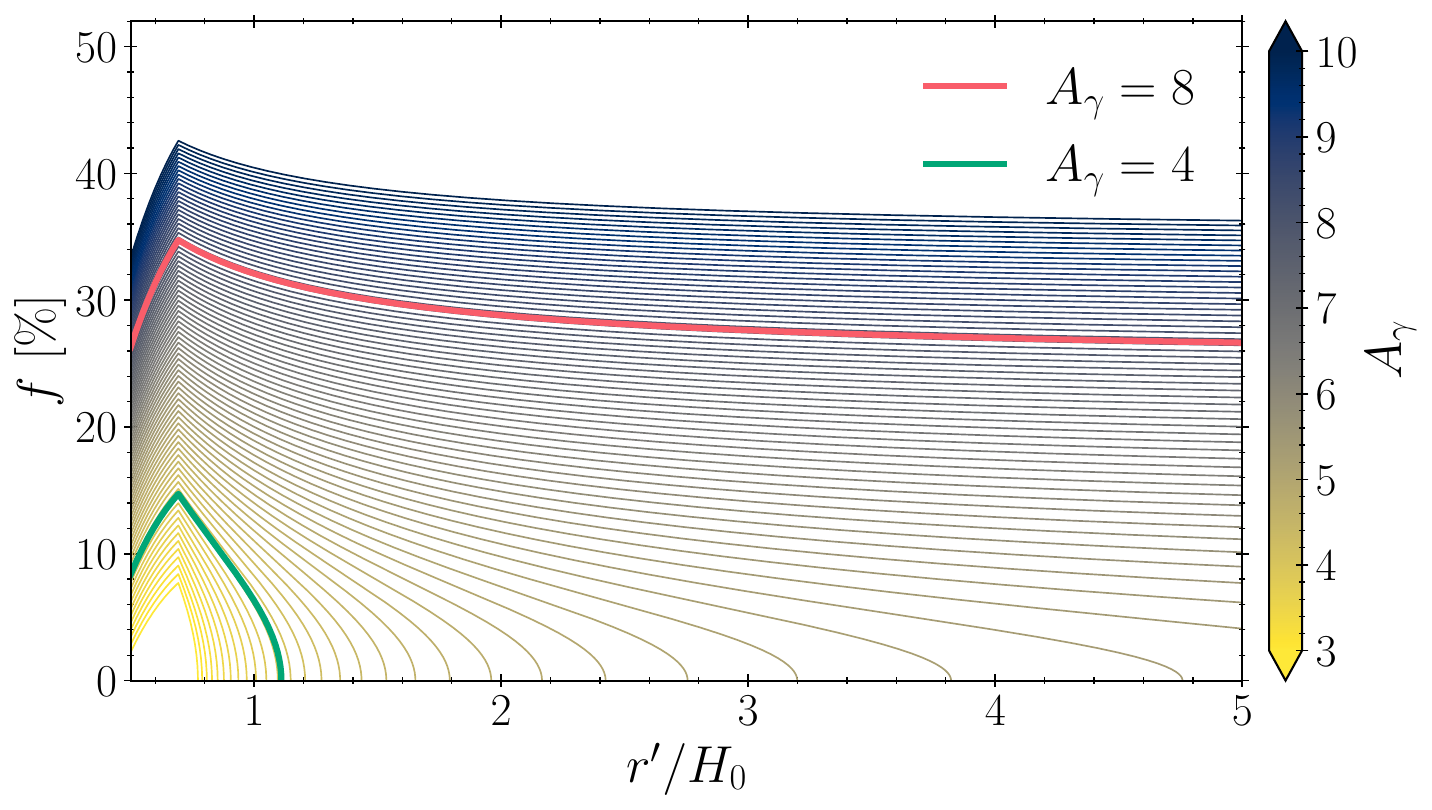}
    \caption{Local kinematic perturbation fraction, $f$, as a function of distance from the planet in units of local scale height. Each curve represents a different wind strength parameter, $A_{\gamma}$, calculated using Equation~\eqref{eq:f_from_Agamma}. The simulation cases for $A_{\gamma} = 4$ and $A_{\gamma} = 8$ are highlighted in green and red, respectively.}
    \label{fig:f_from_Agamma_detailed}
\end{figure}

To facilitate interpretation, we measure the distance from the planet, $r'$, in units of the local disk scale height, $H_0$. Figure \ref{fig:f_from_Agamma_detailed} shows $f(r')$ for a range of $A_{\gamma}$ values. The perturbation strength $f$ reaches a maximum at an intermediate distance from the planet and decreases both near the planet and farther into the disk, consistent with physical expectations. Winds that do not meet the escape condition (see Appendix~\ref{section:Strength parameter}; e.g., $A_{\gamma} = 4$) show a rapid decline in dynamical influence. In contrast, a jet-like case (e.g., $A_{\gamma} = 8$) produces a much stronger and more extended perturbation, reaching $f \sim 0.1-0.3$ over several scale heights.

This analysis neglects effects from viscosity and pressure, which would tend to dampen and disperse the perturbation. The $A_{\gamma}=2$ case is not shown as it does not meet the initial condition for outflow ($\Gamma > G_p$), but it would likely produce only a negligible perturbation. We also ignored the collimation factor, $n$, which confines the wind's effect. A higher degree of collimation could lead to a more localized dynamical effect just above the planet, causing a less extensive perturbation in the surrounding region.

We can extend this analysis to estimate the mass-loss rate, $\dot{M}_w$, driven by the wind. Assuming the outflow expands into a solid angle $\Omega_{\text{eff}}$, the mass-loss rate is:

\begin{equation}
    \dot{M}_w(r') = \Omega_{\text{eff}} r'^2 \rho(r') v_w(r'),
\end{equation}
where $\rho(r')$ is the local gas density. For our chosen collimation ($n=8$), the effective solid angle is $\Omega_{\text{eff}} = 4\pi/9$. Substituting $v_w = f v_k$ and using the value of $f$ from Eq.~\eqref{eq:f_from_Agamma}, we can estimate the mass-loss rate,

\begin{equation}
    \dot{M}_w(r') \approx \frac{4\pi r'^2}{9} \rho(r') f(r') v_k.
    \label{eq:Mdot_estimation}
\end{equation}

\begin{figure}[!ht]
    \centering
    \includegraphics[width=1\columnwidth]{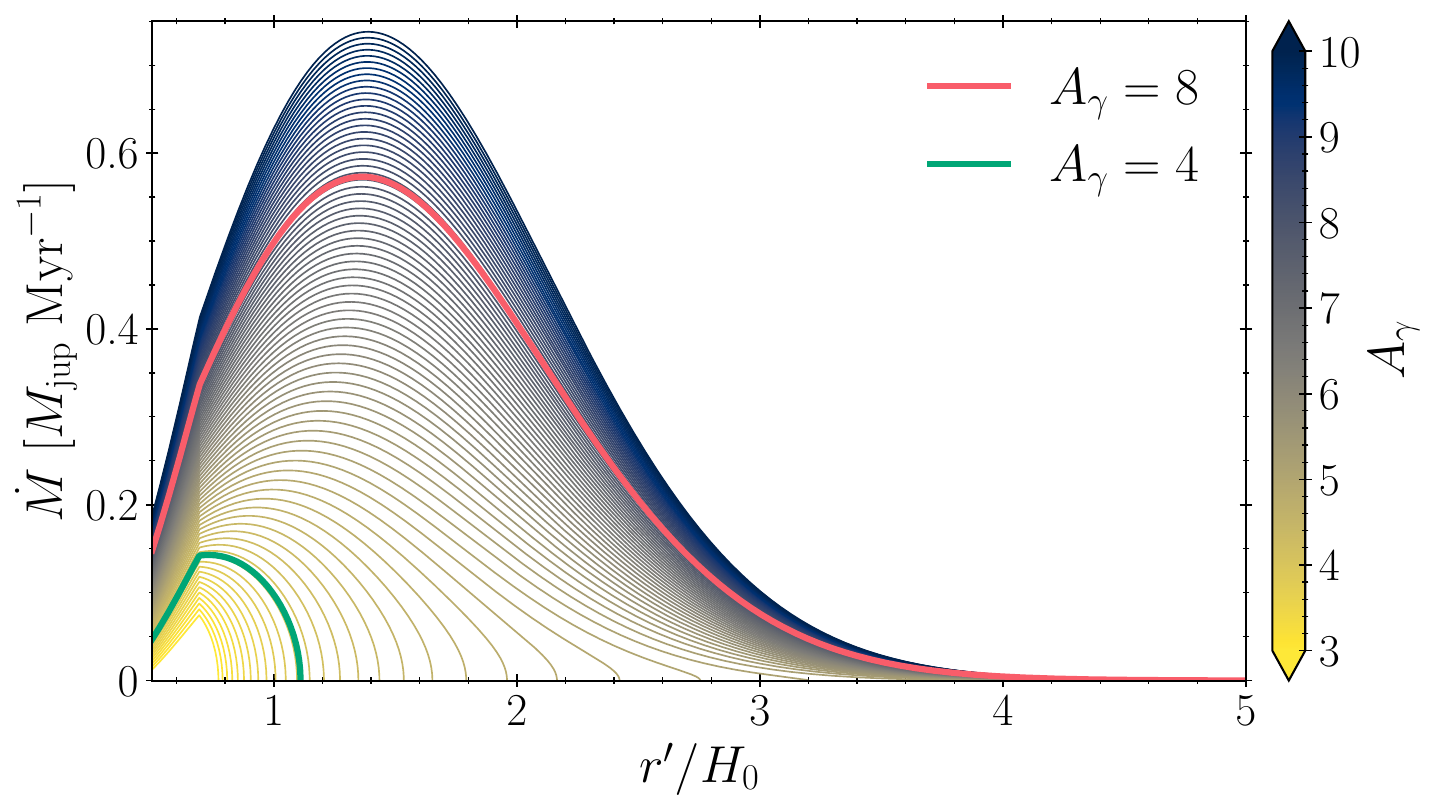}
    \caption{Estimated mass-loss rate $\dot{M}_w$ as a function of distance from the planet (in units of local scale height) for different $A_{\gamma}$ values. The estimate assumes a volumetric density derived from a midplane surface density of $\Sigma_0 = 40\,\mathrm{g\,cm^{-2}}$ at $R_0=10\,\mathrm{AU}$.}
    \label{fig:Mdot_from_Agamma_detailed}
\end{figure}

Figure \ref{fig:Mdot_from_Agamma_detailed} shows the estimated mass-loss rates for our adopted disk parameters. The rates are physically plausible, 
falling within the expected order of magnitude for realistic scenarios, and indicate that the outflow is most effective at driving mass loss in the region where the kinematic perturbation $f$ is strongest. It is important to note that these are instantaneous estimates, corresponding to the initial snapshot. Inside the simulation, the outflow would deplete the local gas density, reducing the mass-loss rate over time.

\subsection{Implementation of planetary accretion} \label{section:AccretionImplementation}

We aim to evaluate the impact the wind has on the planetary accretion process and to measure whether there is a correlation between the accreted mass and the wind strength (mass-loss ratio) as is the case at larger astrophysical scales \citep[e.g.,][]{accretion_frank2002}. Since \texttt{FARGO3D} does not incorporate accretion by default, we modify the code to include this mechanism in our setup.

Previous grid-based studies often incorporate accretion to drive or halt planetary migration via momentum transfer from the accreted material to the planet \citep[e.g.,][]{DePaula2019_acc_fargo3d_gas, Schulik2019_acc_Range1, Nelson2023_acc_fargo3d_gas}. However, as this study focuses on local gas dynamics rather than planetary migration or spin, we employ a simplified accretion model. Here, we neither account for angular momentum transfer nor increase the planetary mass over time, which avoids conflicts with the estimations in Sect.~\ref{section:physical_interpretation_of_Ag}. Consequently, while accretion in our simulations acts as a physical mass sink by removing gas, we use it primarily as a diagnostic tool to compare rates across different setups. The planet's mass and momentum are held fixed and are not updated by the accreted gas, although the gas itself is removed from the computational domain following Eq.\,\eqref{eq:acc-def}.

Our accretion prescription is applied within a sphere of radius $r_{\text{acc}}$ centered on the planet. Within this region, the local gas density $\rho$ evolves according to the differential equation

\begin{equation}
    \frac{\partial\rho}{\partial t} = -\frac{\kappa}{\tau_p} \rho(t),
\end{equation}
which results in an exponential decay of the density, controlled by the accretion efficiency parameter $\kappa$. Here, $\tau_p$ represents the planet's orbital period, and $\kappa$ determines the fraction of mass removed per orbit, which can be interpreted as the fraction of the original gas content that each cell loses after a complete planetary orbit. Numerically, this prescription reads

\begin{equation}
    \rho_{n+1} = \left( \frac{2 - \kappa \overline{\Delta t}}{2 + \kappa \overline{\Delta t}} \right) \rho_n.
    \label{eq:acc-def}
\end{equation}

Here, $\rho_n$ and $\rho_{n+1}$ represent the gas density at the current and next simulation time steps, respectively, and $\overline{\Delta t}$ is the time step normalized to the planet's orbital period. To make the accretion process more physically grounded than a simple sink, we explore two separate criteria for accretion to occur within a cell, tested in independent sets of simulations. The first set requires the material to be gravitationally bound to the planet, while the second requires it to have a negative radial velocity (i.e., be inflowing). Specifically, the conditions are:

\begin{align}
        v_r^2 &< \frac{2GM_p}{r}, \label{eq:acc-condition}\\
        v_r &< 0,
\end{align}
respectively, where $r$ is the distance to the planet, and $v_r$ is the radial velocity relative to the planet. It is important to clarify that these conditions are not applied simultaneously; each is tested in a separate suite of simulations. This approach allows us to investigate separately how the wind affects accretion under two distinct physical assumptions: one based on gravitational binding, which prevents the removal of unbound gas, and another based on inflow dynamics, which ensures only material moving toward the planet is accreted.

\subsection{Physical units}

The simulations in this study use a dimensionless, scale-free framework, allowing the results to be applied to a wide range of physical systems. This approach is well-suited for problems with self-similar dynamics, such as planet-disk interactions, and enhances the generality of our findings. All physical quantities are non-dimensionalized using a set of reference units. The normalization of the system is determined by three characteristic reference parameters:

\begin{itemize}
    \item $R_0$: The semi-major axis of the planet, which sets the unit of length.
    \item $\Sigma_0$: The initial disk surface density at the planet's location, setting the unit for surface density.
    \item $M_{\star}$: The mass of the central star, which sets the unit of mass for point-like bodies.
\end{itemize}

The mass unit for the disk is derived from $\Sigma_0$. From these base units, we derive the characteristic velocity and time:

\begin{itemize}
    \item Velocity unit: $V_0 = \sqrt{GM_\star / R_0}$ (the Keplerian velocity at $R_0$).
    \item Time unit: $T_0 = 2\pi \sqrt{R_0^3 / GM_\star}$ (the orbital period at $R_0$).
\end{itemize}

In Section~\ref{section:Strength parameter}, we introduced the dimensionless wind strength parameter $A_{\gamma}$. Its scaling is justified by the characteristic acceleration in the system, which is set by the star's gravity at the planet's location. This is shown by the equivalence:
\begin{equation}
    \left[\frac{R_0}{T_0^2}\right] \propto \left[\frac{GM_{\star}}{R_0^2}\right]
\end{equation}
This ensures that our wind parameter values correspond to physically consistent regimes regardless of the specific choice of units.

\section{Results} \label{section:Results}

In this section, we present the key results from our hydrodynamical simulations. We analyze two sets of models: one with planetary outflows but no accretion, and another with both outflows and accretion. This approach allows us to isolate the wind's impact on the CPD and to quantify its effect on the accretion rate, which we compare against a baseline simulation without a wind. 

Following the methodology outlined in Sect. \ref{sec:wind_parameterization}, we conduct four simulations with wind strength parameters $A_{\gamma} = 0$, 2, 4, and 8, and with collimation parameter $n = 8$. The $A_{\gamma}=0$ case is the baseline, representing a wind-free scenario. For short, we refer to these models as $\Gamma_0$, $\Gamma_2$, $\Gamma_4$, and $\Gamma_8$, respectively. This range of $A_{\gamma}$ values allows us to explore diverse dynamical regimes, from sub-escape to super-escape conditions. All simulations assume a purely polar wind ($\theta_{\text{inc}} = 0$). To provide a concrete physical context for our dimensionless results, we scale them using the representative parameters listed in Table~\ref{tab:physical_params}.

\begin{table}[h]
\caption{Reference physical parameters used to scale the dimensionless simulation results.}
\label{tab:physical_params}
\centering
\begin{tabular}{@{}ll@{}}
\hline
\hline
\textbf{Parameter} & \textbf{Value} \\
\hline
$R_0$ & $10\,\mathrm{au}$ \\
$M_{\star}$ & $2.11\,M_{\odot}$ \\
$\Sigma_0$ & $40\,\mathrm{g\,cm^{-2}}$ \\
$M_p$ & $2.11\,M_{\mathrm{jup}}$ \\
\hline
$V_0$ & $\sim 4.32\,\mathrm{km\,s^{-1}}$ \\
$T_0$ & $\sim 6.88 \times 10^{-4}\,\mathrm{Myr}$ \\
\hline
\end{tabular}
\end{table}

\subsection{Isolated planetary outflow model} \label{sec:isolatedPlanetaryOutflowModel}

\begin{figure*}[p]
    \centering
    \includegraphics[width=0.7\textwidth]{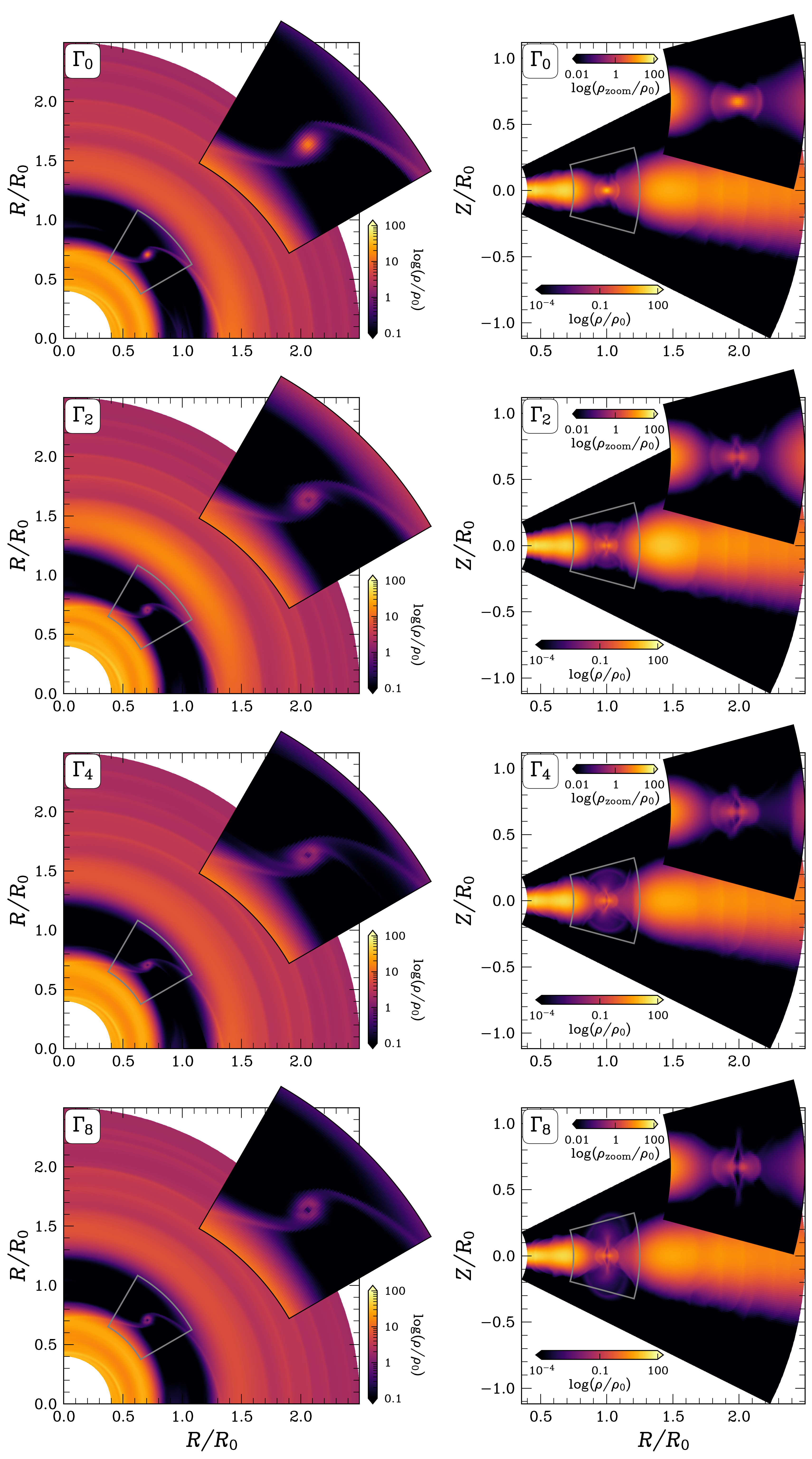}
    \caption{Comparative snapshots of simulations $\Gamma_0$, $\Gamma_2$, $\Gamma_4$, and $\Gamma_8$ (from top to bottom), all taken at 500 planetary orbits. The left column shows the logarithmic gas density in the disk midplane ($R$-$\phi$), with a zoomed-in view centered on the planet; both panels share the same color scale. The right column presents meridional slices ($R$-$Z$) of the logarithmic gas density. The full disk view and the zoomed-in region use different color scales, with the latter denoted by $\rho_{\mathrm{zoom}}$, to highlight both global and local density structures.}
    \label{fig:2dSnapshotGammaNFinal}
\end{figure*}

The simulations ran for 500 orbits at $r=10\,\mathrm{au}$, equivalent to $t\approx 0.01$~Myr considering the scaling of the system, which is enough time to observe the long-term consequences of the wind and the planet's presence. The wind and the planet are both introduced at the beginning of the simulations (0 orbits), so their effects are visible from the start, with no ramp-up in their potentials. We present comparative snapshots of the final stage of all simulations, at $t=500$ orbits in Fig.~\ref{fig:2dSnapshotGammaNFinal}, showing views of the equatorial and meridional planes at the planet's position.

In all cases, we observe the opening of a gap, a direct result of the planet's influence. The overall disk structure does not show noticeable variations due to the wind's presence. In the $\Gamma_2$ case, a lobe of overdensity, more pronounced than in the snapshots of the other models, is visible near the planet at $R/R_0 \sim 1.5$ (see Fig.~\ref{fig:2dSnapshotGammaNFinal}); however, we interpret this as a transient feature of this particular snapshot rather than a stable structure, as similar transient overdensities appear and dissipate intermittently throughout all simulations.

\begin{figure*}[!h]
    \centering
    \includegraphics[width=0.8\textwidth]{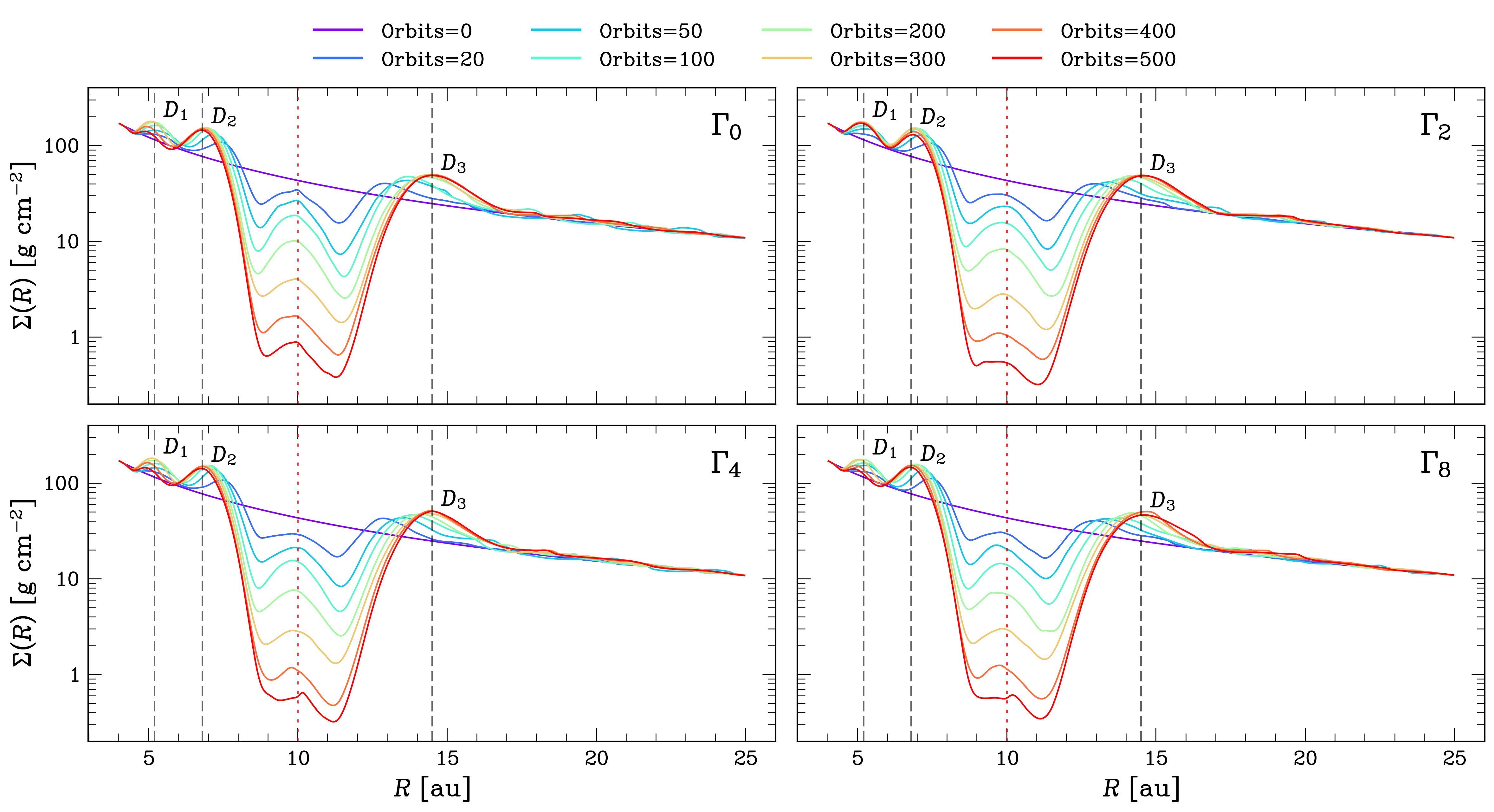}
    \caption{Surface density profile evolution for $\Gamma_0$ (top-left), $\Gamma_2$ (top-right), $\Gamma_4$ (bottom-left), and $\Gamma_8$ (bottom-right). The planet's position at 10~au is marked by a red dashed line. The planet opens a primary gap and excites three overdensities ($D_1, D_2, D_3$). Surface density profiles are obtained by integrating the volumetric density over the polar angle ($\theta$) and then averaging over the azimuthal angle ($\phi$) for each orbit.}
    \label{fig:densityProfile_evolution}
\end{figure*}

\begin{figure*}[!h]
    \centering
    \includegraphics[width=0.9\textwidth]{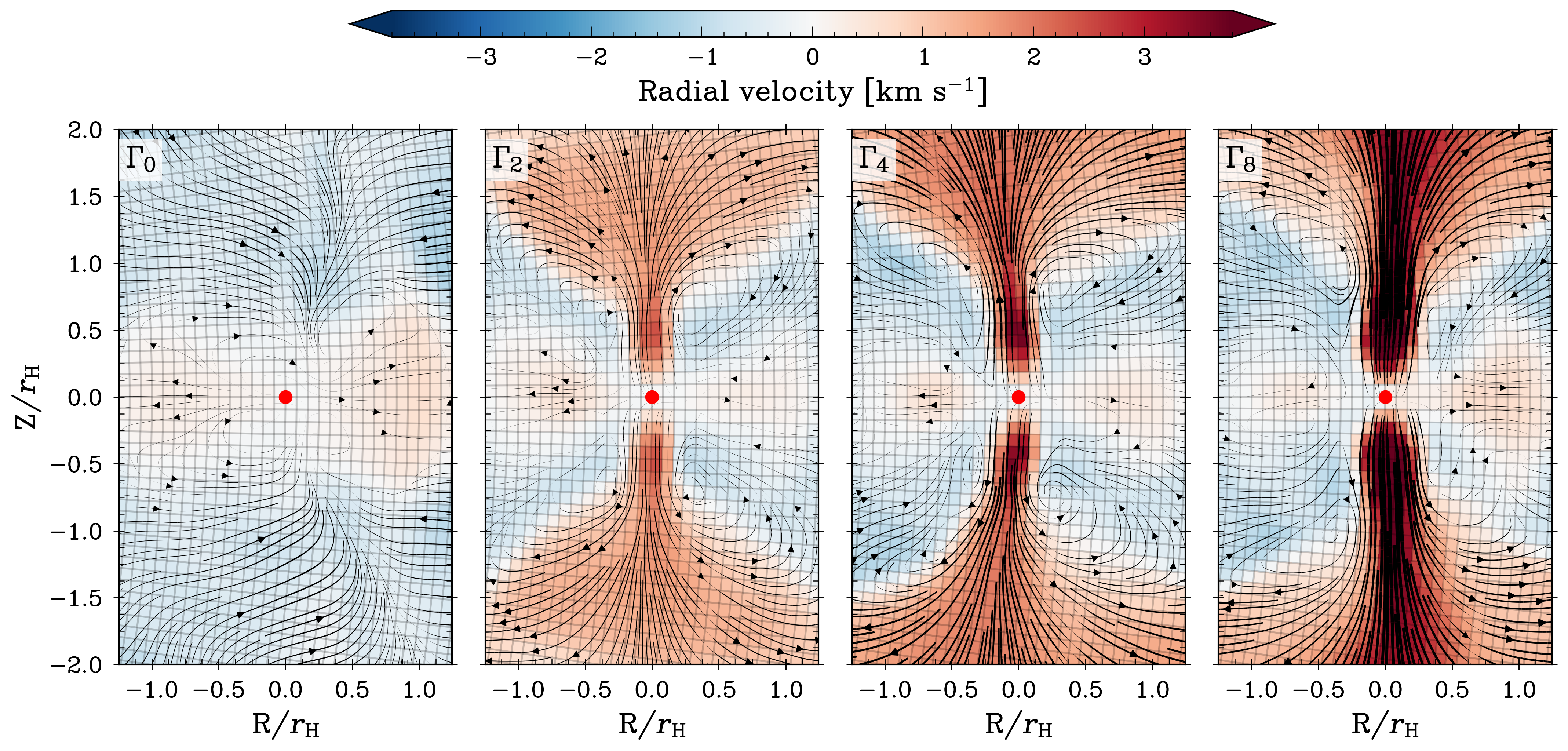}
    \caption{Meridional snapshot at 500 orbits showing the radial velocity relative to the planet for different wind strengths: $\Gamma_0$, $\Gamma_2$, $\Gamma_4$, and $\Gamma_8$.}
    \label{fig:FlowLines}
\end{figure*}

To test whether the wind could cause structural perturbations that might be hidden by a massive planet, we also ran simulations with a lower planetary mass. However, no significant structural changes were found for the purely meridional outflows studied here, indicating that this particular wind configuration is incapable of generating large-scale disk perturbations. This is supported by Fig.~\ref{fig:densityProfile_evolution}, where the density profiles show no significant variations. Nevertheless, a closer look at the disk's ring and gap structure reveals some subtle perturbations.

In the baseline $\Gamma_0$ model, the planet opens a primary gap and excites three overdensities, which we denote as $D_1$, $D_2$, and $D_3$, located at approximately 5.2, 6.8, and 14.5~au, respectively. The innermost overdensity, $D_1$, is a transient feature that vanishes after approximately 300 orbits for all models except $\Gamma_2$, where it remains stable after 500 orbits (see Fig.~\ref{fig:densityProfile_evolution}). The second overdensity, $D_2$, shows a different behavior, remaining stable in the $\Gamma_0, \Gamma_4,$ and $\Gamma_8$ models, but exhibiting a significant density decrease over time in the $\Gamma_2$ simulation. The outermost overdensity, $D_3$, remains stable across all simulations. While we cannot entirely rule out influence from the inner boundary conditions, these differences suggest that even a weak wind ($\Gamma_2$) can introduce subtle, large-scale redistributions of material, a point that is intriguing given the wind's localized nature. While the depth of the gap is greater in the wind-active cases, the differences are subtle, at about 20\%, which is negligible in the context of this investigation.

Figure 5 shows that gap formation remains ongoing at the end of our simulations. Since the gap width and depth evolve on viscous timescales \citep{Kanagawa2017}, reaching a quasi-steady configuration would require integration times much longer than ours. More importantly, since we consider a low-viscosity disk and planets at large orbital radii, this timescale would likely exceed the lifetime of the protoplanetary disk. Although a deeper gap could quantitatively affect mass transport efficiency, the qualitative presence and structure of the planet-induced gap are already well established.

Turning our focus to the planet's surroundings, in Fig.~\ref{fig:2dSnapshotGammaNFinal}, the outflow leaves a physical trace, displacing material in the vicinity of the planet. From the equatorial view, we see that the amount of material very close to the planet diminishes drastically due to the action of the wind, creating a low-density cavity at the planet's position. This effect becomes progressively more pronounced with increasing wind strength, indicating a direct correlation between wind intensity and gas removal efficiency around the planet.

In the meridional plane, the wind's impact is further evident through two key features that shape the dynamics and mass distribution around the planet. First, we observe the formation of distinct low-density lobes aligned with the outflow direction, forming bipolar cavities produced by the rapid evacuation of gas along the polar axis. These regions reflect a persistent reduction of material within the planet's Hill sphere, an effect that becomes more pronounced as the wind strength increases. Simultaneously, Fig.~\ref{fig:2dSnapshotGammaNFinal} shows the presence of an overdense structure surrounding the wind, which is particularly prominent at early epochs. This structure, referred to as a "blob" (see Sect.~\ref{section:Strength parameter}), forms from material displaced by the outflow but still gravitationally bound to the planet. Lacking the energy to escape, the  gas accumulates along the outer contours of the wind and eventually falls back, producing a transient overdensity that surrounds the outflow. This redistribution of material is further illustrated in Fig.~\ref{fig:FlowLines}, where we plot the radial velocity relative to the planet, defined as

\begin{equation}
    v_r = \bm{v} \cdot \hat{r}^{\prime \prime},
\end{equation}
where $\hat{r}^{\prime \prime}$ points outward from the planet. This figure highlights the evolution of gas dynamics around the planet, and the impact of the wind's presence and strength.

In the absence of a wind, $\Gamma_0$, we observe the standard polar accretion configuration dominating, with mass flowing primarily along the polar axes toward the planet, as expected in young planetary systems. However, this polar inflow is rapidly disrupted once the wind is activated, even at the lowest intensity ($\Gamma_2$). Instead, accretion shifts toward the midplane, indicating that the wind efficiently suppresses vertical inflow. This shift in accretion geometry is robust across all wind-influenced cases ($\Gamma_2$, $\Gamma_4$, $\Gamma_8$), emphasizing the sensitivity of polar accretion to meridional outflows. These results are consistent with the findings of \citet{Machida2006}, who similarly reported that strong vertical flows can significantly alter the accretion pattern onto the planet.

\begin{figure}[!h]
    \centering
    \includegraphics[width=1\columnwidth]{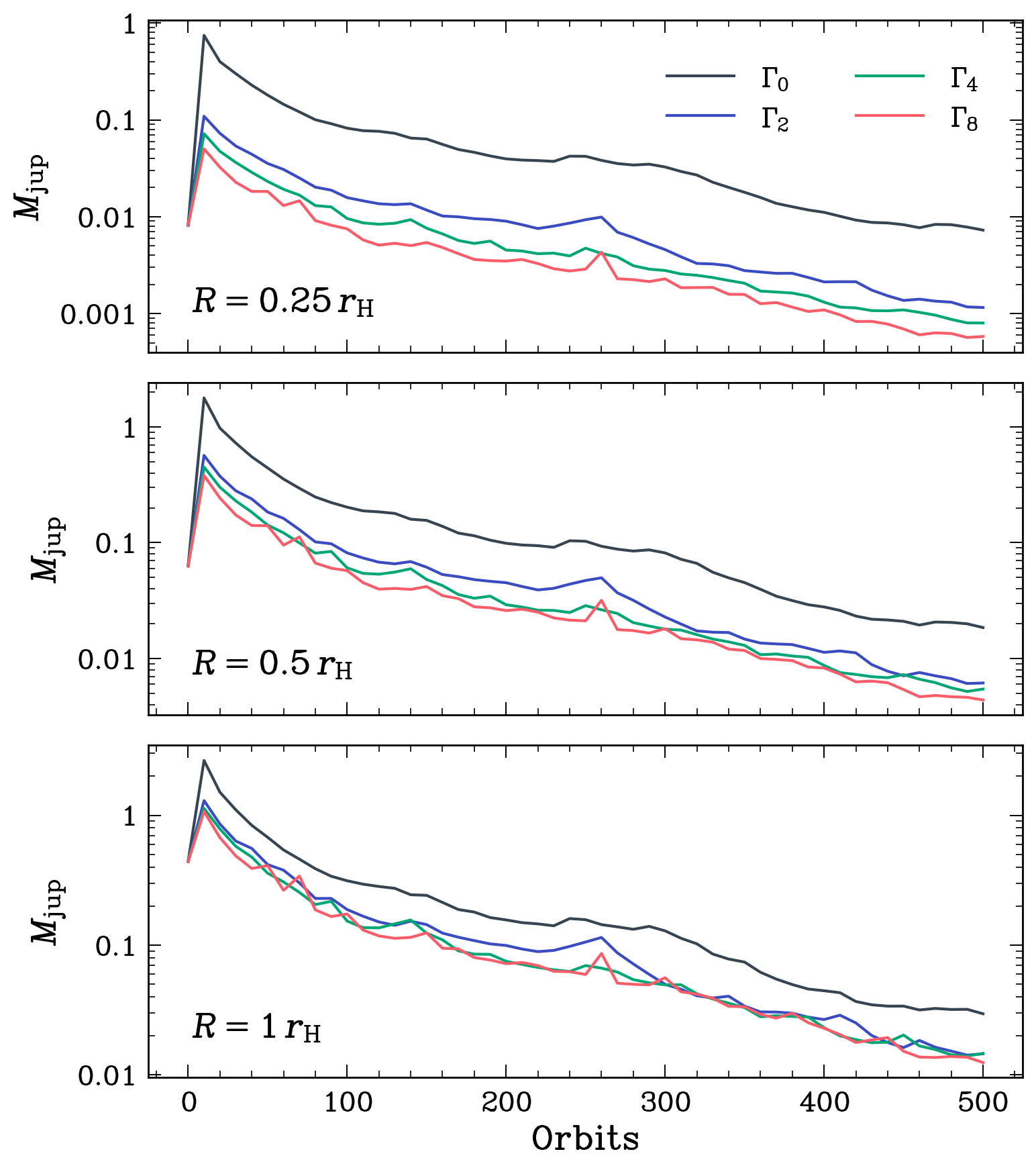}
    \caption{Mass within the radii $0.25 r_{\mathrm{H}}$, $0.5 r_{\mathrm{H}}$, and $1 r_{\mathrm{H}}$ shown in the top, middle, and bottom panels, respectively, over time for simulations $\Gamma_0$ in black, $\Gamma_2$ in blue, $\Gamma_4$ in green, and $\Gamma_8$ in red.}
    \label{fig:massShell_gammaN_log}
\end{figure}

We track the evolution of the mass enclosed within radii of 0.25~$r_{\text{H}}$, 0.5~$r_{\text{H}}$, and 1~$r_{\text{H}}$ around the planet (see Fig.~\ref{fig:massShell_gammaN_log}) to evaluate the wind's influence as a gas removal mechanism. Initially, a rapid increase in mass occurs within the Hill sphere during the first $\sim10$ orbits. However, the presence and strength of the wind significantly affect the mass that can be retained by the planet, particularly in the inner regions ($0.25 r_{\mathrm{H}}$). Even the weakest wind ($\Gamma_2$) results in a difference of nearly an order of magnitude in accumulated mass compared to the wind-free case. After this point, a decline in mass is observed across all simulations, driven by the formation of the horseshoe region, which gradually clears material from the planet's vicinity. This mass loss occurs at a similar rate in all cases, as indicated by the similar slopes, suggesting that the primary driver of the long-term depletion is not a persistent effect of the outflow.

To further examine this behavior, we analyze the mass flux in the planet's reference frame, $\mathbf{S''}$. The radial mass flux is defined as:

\begin{equation}
    \dot{M} \equiv \rho \ \bm{v} \cdot \hat{r}^{\prime \prime}
    \label{eq:massFluxDef}
\end{equation}
where $\rho$ is the instantaneous volumetric density, $\bm{v}$ is the instantaneous velocity, and $\hat{r}^{\prime \prime}$ is the outward unit vector relative to the planet. By convention, a positive flux ($\dot{M} > 0$) represents material flowing outward (outflow), while a negative flux corresponds to material flowing inward (inflow).

\begin{figure}[!h]
    \centering
    \includegraphics[width=1\columnwidth]{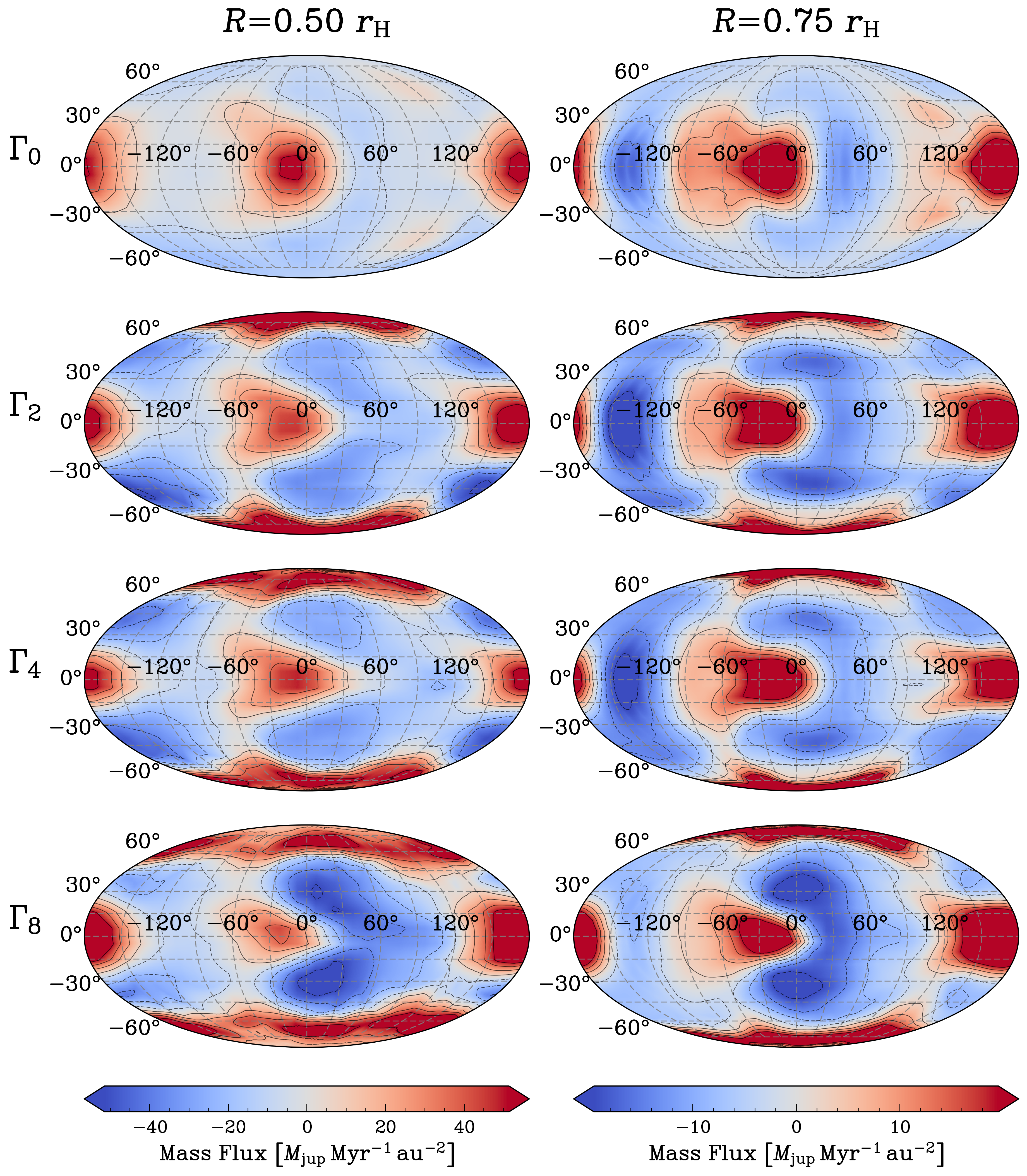}
    \caption{Mass flux (see Equation~\eqref{eq:massFluxDef}) projected at radii of $0.5\,r_\mathrm{H}$ and $0.75\,r_\mathrm{H}$ (left and right columns, respectively) for simulations $\Gamma_0$, $\Gamma_2$, $\Gamma_4$, and $\Gamma_8$ (top to bottom rows) at 100 planetary orbits. The color scale represents the mass flux density in units of $M_{\mathrm{jup}}\,\mathrm{Myr}^{-1}\,\mathrm{au}^{-2}$. Positive values (red) indicate outflow, while negative values (blue) indicate inflow.}
    \label{fig:shellFluxMaps}
\end{figure}

Two-dimensional projections of the mass flux at selected radial distances from the planet provide a visualization of the accretion pattern \citep[e.g.,][]{Gressel2013GLOBAL, Tanigawa2012_distributionAccretion}. These maps are presented in Fig.~\ref{fig:shellFluxMaps}, with projections at $0.5\,r_\mathrm{H}$ and $0.75\,r_{\mathrm{H}}$ at 100 orbits; projections at lower radii are avoided since they are at the limit of our grid resolution. As expected, our results show variability over time, which we have compiled into as an online movie covering the 500 orbits. In all cases, we observe the appearance of strong lobes of inflow and outflow at the equatorial regions ($\mathrm{Latitude} \sim 0^{\circ}$), which aligns with the spiral arm positions and dynamics. For the wind-active simulations, the outflow component in the poles is evident and becomes more pronounced with stronger outflows.

\begin{figure}[!h]
    \centering
    \includegraphics[width=1\columnwidth]{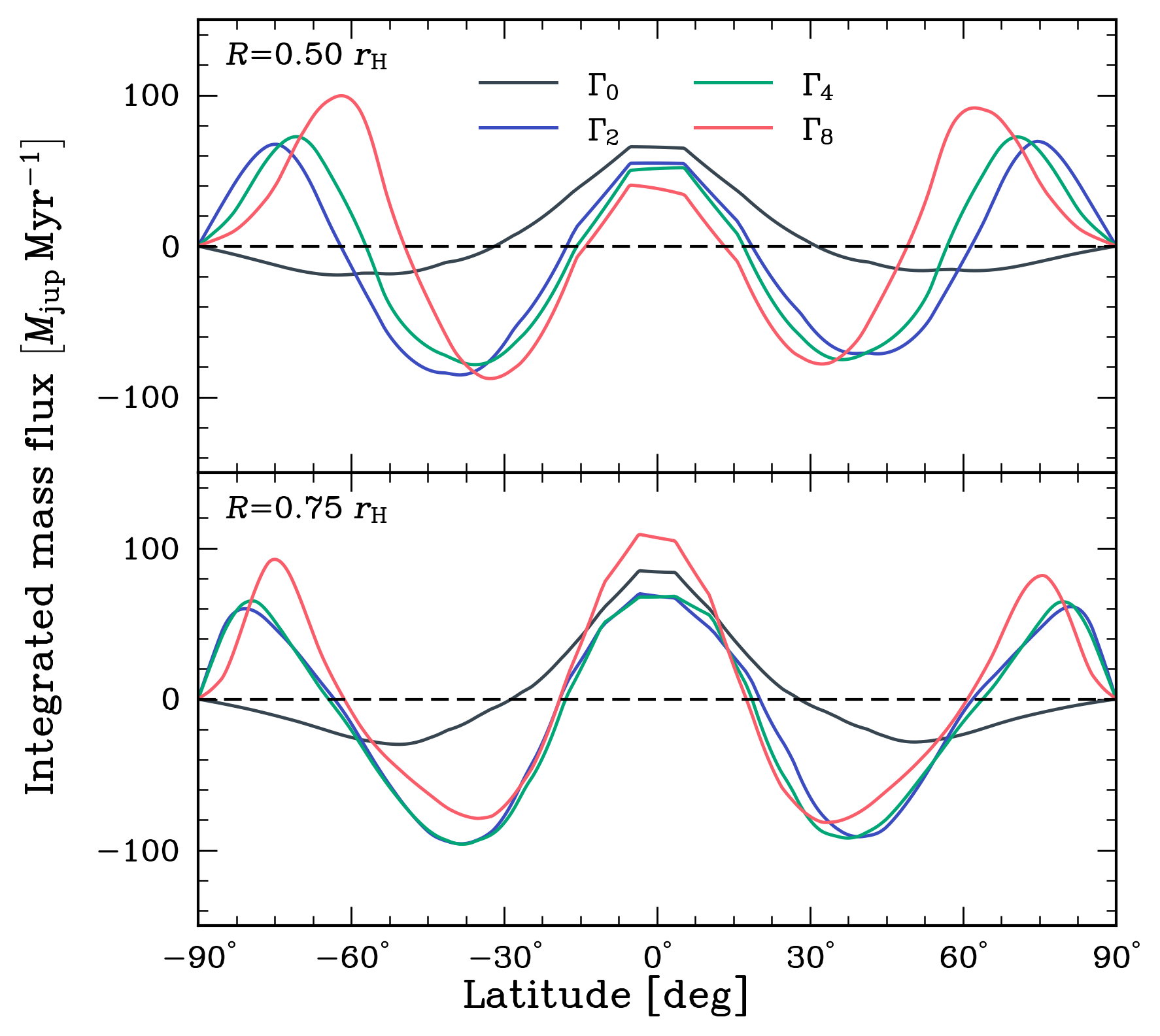}
    \caption{Azimuthally integrated mass fluxes as a function of latitude for all simulations shown in Figure~\ref{fig:shellFluxMaps}. Positive flux indicates outflow, while negative flux indicates inflow.}
    \label{fig:avgf_by_lat}
\end{figure}

A comparison of the net wind accretion pattern as a function of co-latitude can be carried out by extracting the  azimuthally averaged mass flux. Figure~\ref{fig:avgf_by_lat}  shows signatures of mass depletion in the midplane, an effect that appears more pronounced for wind-active cases. The wind also imposes strong outflow components in the poles. Although weaker winds appear to lead to greater mass outflow, this is a consequence of the lower density in the polar regions for these cases, where stronger winds have already ejected most of the mass. Consistent with Figure~\ref{fig:FlowLines}, we observe an isotropic inflow component in the planet's polar regions, in agreement with a polar accretion scenario for $\Gamma_0$. However, when the wind is activated, this inflow shifts toward the contours of the outflow, which are predominant at latitudes of $\pm 30-50^{\circ}$ and dominate the mass inflow. These results support the idea of a preferential redirection of material toward the midplane in response to the outflows. It is worth pointing out that this effect is consistent even for the weakest wind, $\Gamma_2$.

\subsection{Accretion-enabled planetary outflow model} \label{sec:AccretionResults}

The motivation to include accretion comes from the results shown in Fig.~\ref{fig:FlowLines} and Fig.~\ref{fig:shellFluxMaps}, where the flow lines near the planet and the appearance of dense blobs suggest that wind-driven dynamics could affect how gas is distributed and delivered onto the planet. To investigate this, we adopt the accretion prescriptions defined in Sect.~\ref{section:AccretionImplementation}, which depend on two parameters: $\kappa$, controlling the fraction of mass accreted per time step, and $r_{\mathrm{acc}}$, defining the radial extent of the accretion region. We test both conditions presented in Sect.~\ref{section:AccretionImplementation} in separate sets of simulations. Studying these accretion conditions separately allows us to distinguish the efficiency of each and determine which dominates during different periods. For practical and detailed comparisons, we avoid drawing conclusions from the first 5--10 orbits, as the system is still settling to the presence of the planet and the accretion prescription. Results beyond this point are more informative and provide better insight into the underlying dynamics.

\begin{figure*}[h!]
    \centering
    \includegraphics[width=0.85\textwidth]{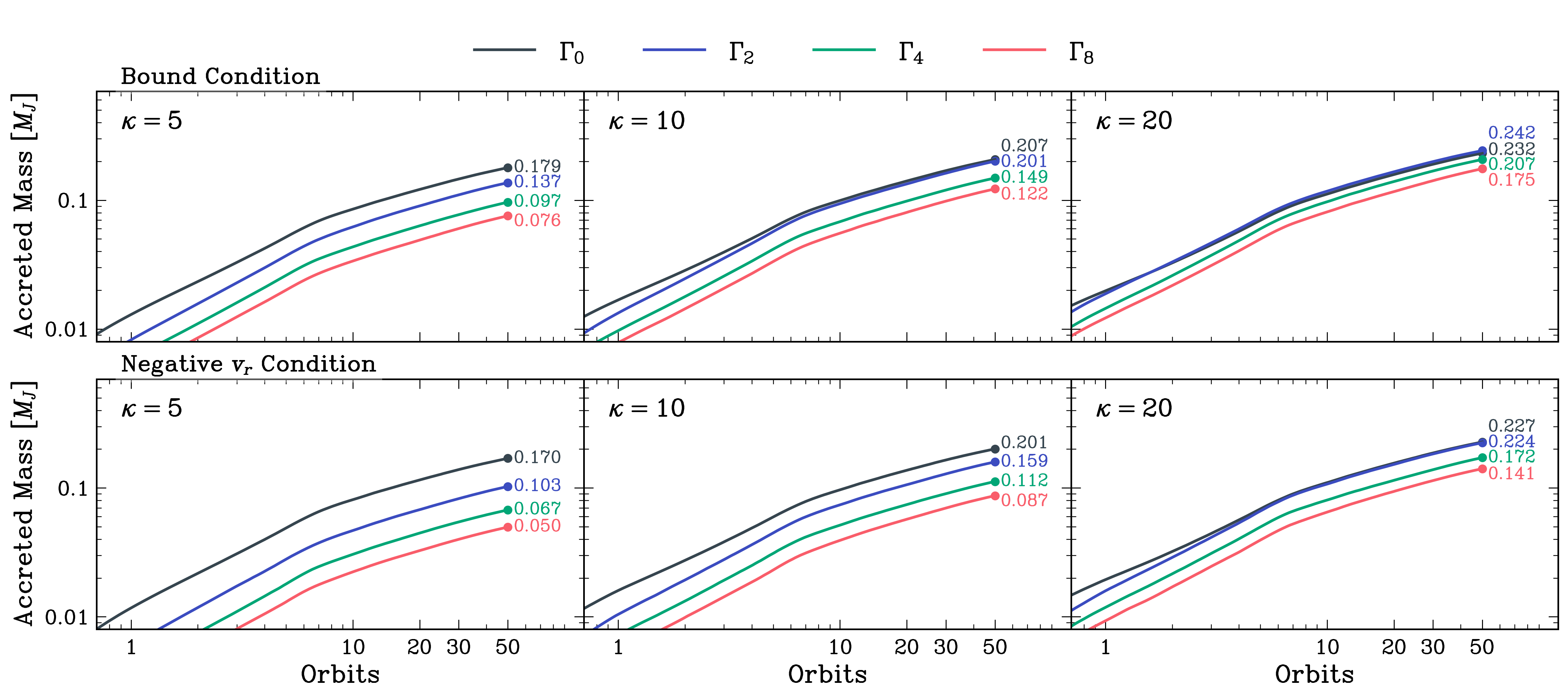}
    \caption{Total accreted mass over time for different accretion efficiencies ($\kappa=5, 10, 20$) and physical conditions. Black, blue, green, and red lines correspond to $\Gamma_0,\, \Gamma_2,\, \Gamma_4,\, \mathrm{and} \, \Gamma_8$, respectively. Numbers in each panel show the total accreted mass at 50 orbits.}
    \label{fig:accretion_analysis}
\end{figure*}

We limited this process to a compact region near the planet by setting the accretion radius to $r_{\mathrm{acc}} = r_{\mathrm{H}}/4$, which is at the limit of our grid resolution. This choice is intentionally conservative, as it isolates local effects and acknowledges that our model cannot properly resolve the gas dynamics at smaller distances. Furthermore, in nature, accretion is expected to occur at the scale of the planetary radius. Since the amount of accreted material naturally scales with the accretion radius, selecting a value at our resolution limit facilitates convergence for future studies with higher resolution that can probe scales closer to the planet. 

Within this numerical framework, we explored $\kappa = \{5, 10, 20\}$ for four wind strengths ($\Gamma_0, \Gamma_2, \Gamma_4, \Gamma_8$), with the two physical conditions presented in Sect.~\ref{section:AccretionImplementation} treated separately, for a total of 24 simulations. Each simulation was run for 50 orbits ($\sim 0.0016$~Myr), which, while only 10\% of the integration time of the models in Sect.~\ref{sec:isolatedPlanetaryOutflowModel}, is sufficient to capture the early-time evolution and the direct impact of the wind. Accretion was activated at the beginning of the simulations, with no ramp-up period.

As discussed in Sect.~\ref{section:AccretionImplementation}, this approach fixes the planet's mass and position and neglects angular momentum exchange with the accreted material, which constrains the interpretation of the resulting accretion rates. Figure~\ref{fig:accretion_analysis} shows the total accreted mass over time for the two accretion conditions tested. Although a quasi-steady state is not reached within the 50-orbit runtime, the early-time behavior provides a meaningful comparison of the wind's initial influence on accretion.

The total accreted mass over time is broadly similar for both accretion criteria, although the bound material condition is consistently more efficient. For the baseline model ($\Gamma_0$), the difference between the two criteria is less than 10\%. However, for wind-active cases, this difference becomes more significant, ranging from 10\% to 50\%, with the largest discrepancies observed for stronger winds. This suggests that while both criteria capture the accretion process, the wind's presence accentuates the distinction between what is merely inflowing and what is gravitationally captured.

In general, our results show that planetary winds have a negative impact on the total accreted mass. As shown in Fig.~\ref{fig:accretion_analysis}, stronger winds (higher $A_\gamma$) lead to lower accreted mass, a trend that aligns with the overall mass depletion observed within the planet's Hill sphere (Fig.~\ref{fig:massShell_gammaN_log}). This reinforces the idea that winds limit the feeding of material into the planetary region.

This trend, however, is modulated by the accretion efficiency, $\kappa$. As $\kappa$ increases, the differences in accreted mass between the wind-active models and the baseline model diminish. For instance, the accretion rate for the weak-wind model ($\Gamma_2$) rapidly approaches that of the no-wind model ($\Gamma_0$) as the accretion process becomes more efficient. This convergence can be explained by the competition between accretion and wind-driven redistribution: when $\kappa$ is high, the removal of available gas via accretion outpaces its redistribution by the wind. As a result, the wind has less time to disperse material before it is accreted.

An interesting exception to the general trend is observed for the $\Gamma_2$ model under the bound material condition with high efficiency ($\kappa=20$). In this case, the total accreted mass is slightly higher ($\sim4.3\%$) than in the baseline $\Gamma_0$ model. The accretion curves for both models become very similar after just one orbit, suggesting that a weak, persistent outflow might not always hinder accretion and could, under certain conditions, slightly enhance it by modifying local flow patterns in a way that favors capture of bound material.

For a more detailed analysis of the spatial distribution of the accreted material, we project the accretion rate at $0.15\,r_{\mathrm{H}}$ from the planet and overlay the mass flux. These maps are presented in Appendix~\ref{section:appendix_accretionPathways} (see Figs.~\ref{fig:projectionMap_macc_kappa05}, \ref{fig:projectionMap_macc_kappa10}, and \ref{fig:projectionMap_macc_kappa20}). The asymmetries observed in these projection maps result from the limited grid resolution in this region. In all cases, the accretion regions coincide with contours of inflowing material. The baseline model ($\Gamma_0$) exhibits accretion from almost all directions. Conversely, in wind-active models, accretion is dominated by flows from the midplane. This accretion window narrows as the outflow strength increases, consistent with the conclusion that planetary outflows shift accretion pathways toward equatorial regions. This spatial redistribution offers an explanation for the case where $\Gamma_2$ accretes more mass than $\Gamma_0$. Since accretion in the wind-active scenario is restricted to the midplane, the higher total accretion rate implies that the wind enhances the mass supply from this region significantly more than what is observed in the nominal case. However, the overall accretion remains comparable, as the total accreted mass curve follows a shape very similar to that of $\Gamma_0$. Thus, even though the wind locally enhances accretion from the midplane, the global accreted mass remains largely unaffected.

\section{Discussion and conclusions} \label{section:Conclusions}

Our simulations explore the influence of planetary outflows on gas dynamics and accretion processes in the vicinity of embedded planets. We formulated an analytical outflow prescription and considered a range of wind strengths, from a no-wind case to progressively stronger meridional outflows up to jet-like ejections, to understand their impact on disk morphology, disk kinematics, and flow structures. Below, we discuss the key findings, limitations, and implications of our results.

\subsection{Wind-driven gas dynamics}

The study of mass accretion onto growing planets is challenged by the current inconclusive understanding of which direction dominates inflow in planet-forming systems. Hydrodynamical models typically find a strong component of mass inflow from the polar regions, and analytical predictions often suggest infalls of supersonic material, supporting a polar-dominated accretion picture \citep[e.g.,][postfix]{Montesinos2025_ionizedEnvelopesAroundProtoplanets}. However, numerical simulations tend to be limited by resolution, rarely extending beyond a fraction of the Hill sphere ($r_{\mathrm{H}}$) and are often unable to properly resolve the dynamics in the innermost planetary vicinity, including the CPD. High-resolution analyses, such as \cite{Tanigawa2012_distributionAccretion}, have approached closer to the planet, finding important accretion components at about 60~$^{\circ}$ from the midplane. This underscores the necessity of investigating the dynamics in these close-in regions to properly resolve them.

Our simulations reveal that in the absence of wind, mass inflow occurs from virtually all directions, with significant contributions from both polar and midplane regions. However, our mass flux maps (see Fig.~\ref{fig:shellFluxMaps}) show that even the weakest simulated outflow, $\Gamma_2$, is sufficient to disrupt this process entirely. The flows are redirected toward the equatorial plane, with mass flux peaking near 30~$^{\circ}$ from the midplane. This trend is consistently observed across all wind-active simulations, indicating that the accretion geometry is highly sensitive to planetary winds in general, not only polar outflows, as similar results are observed when the inclination of the primary outflow direction, $\theta_{\mathrm{inc}}$, is considered. Since the strength of $\Gamma_2$ is weaker than the planet's gravity throughout the solved region, we attribute the emergence of this phenomenon not to the outflow's natural strength, but rather to the intrinsic alterations it introduces to the planetary acceleration field, hindering the inflow path through the poles and instead facilitating inflow through the outflow's periphery where its effect becomes minimal, leading to the observed redirection. For the stronger cases of $\Gamma_4$ and $\Gamma_8$, the same conclusions can be drawn for the weaker contours, with the addition of a strong outflow component in the wind's primary direction. We expect variations in the primary inflow angle with the $n$ value, with lower $n$ leading to broader winds and inflow angles closer to the midplane. It is important to note, however, that the particular wind configuration studied here, which is purely meridional, showed no significant structural perturbations on the global disk. Even in exploratory simulations with inclined and spinning winds, no major large-scale perturbations were observed. We attribute this to two main factors. First, the wind's impact is confined by the imposed cutoff radius ($r_{\mathrm{H}}/2$), limiting its direct influence to the planet's immediate vicinity. Second, since the wind acts as an acceleration, its ability to perturb the disk is proportional to the local gas mass. As the wind itself causes significant mass depletion around the planet, it creates a self-limiting feedback loop: a stronger wind evacuates its immediate environment more effectively, thereby reducing its own capacity to exert force on the larger disk structure. A noteworthy and unresolved finding from our 500-orbit simulations is the anomalous behavior of the weakest-wind model, $\Gamma_2$ (see Fig.~\ref{fig:densityProfile_evolution}). This model alone showed significant, large-scale structural perturbations, such as the persistence of the overdensity at $\sim 5.2$~au ($D_1$) and a shift in the unperturbed density region, features not seen in the wind-free or stronger-wind cases. This is counterintuitive, as one might expect stronger winds to have a greater structural impact. We did not find a clear physical explanation for this, though we speculate it could be related to a feedback loop that is only effective for a weak, persistent outflow and is otherwise suppressed or overpowered by the stronger winds. This intriguing result warrants further investigation with dedicated simulations to determine if it is a robust physical phenomenon or an artifact of our model's specific parameters or boundary conditions.

\subsection{Mass depletion and its effect on planetary growth}

The primary consequence of the wind-driven flow redirection is a significant depletion of the gas reservoir within the planet's vicinity (see Fig.~\ref{fig:massShell_gammaN_log}). This effect is more pronounced for stronger winds but remains significant even for the weakest outflow, with a mass difference of almost an order of magnitude in the inner regions ($r < 0.25\,r_{\mathrm{H}}$). The wind's impact is most dramatic during the initial orbits, where it severely limits the amount of gas the planet can capture, effectively reducing the mass reservoir available for planetary growth.

Over longer timescales, we observe a steady mass decline in all models, including the wind-free case. We attribute this secular depletion to the natural dynamics of the horseshoe region, where material is exchanged across the planet's orbit, leading to a gradual leakage of gas from the Hill sphere. Planetary winds compound this effect; they do not necessarily increase the mass outflow rate over long periods but rather act as a barrier, restricting the net inflow of material to the planet. As a hydrodynamical phenomenon, this wind-driven mass depletion is expected to be most effective for gas and small, well-coupled dust, suggesting that such outflows could be a key mechanism for slowing and regulating the growth of gas giants.

Our accretion-enabled simulations (Sect.~\ref{sec:AccretionResults}) largely support this interpretation. Generally, the presence of an outflow leads to lower accretion rates compared to the wind-free scenario, confirming that winds diminish the material available for planetary growth. This highlights planetary winds as a potential mechanism for halting planetary growth at early stages, preventing the overgrowth of compact objects in some cases (see Figs.~\ref{fig:massShell_gammaN_log} and \ref{fig:accretion_analysis}). However, we observe an exception for weak winds ($\Gamma_2$) under high accretion efficiency, where accretion rates can match or slightly exceed the baseline. This suggests that while winds deplete the global envelope, they can locally enhance accretion flux from the midplane by redirecting flows, partially compensating for the loss of polar inflow. However, our parametric model does not enforce a physical link between the accretion rate and the outflow strength. In real astrophysical systems, such as those around young stars or black holes, a tight correlation is often observed where a fraction of the accreted material \citep[typically from 0.01 to 0.1, e.g.,][]{Fang2018_molecuarAnalisysForAccretionOutflowCorrelation,Rota2025_correlationAccretionOutflow, Mestici2024_correlationAccretionOutflowAGNs} is ejected in an outflow. While our simulations show that gas dynamics alone are not sufficient to produce such a correlation, we do not rule out its existence. A self-consistent model that links the outflow to a physical launching mechanism (e.g., accretion luminosity or magnetic fields) would be required to explore this relationship in the planetary context, but this is beyond the scope of the present work.

\subsection{The nature of circumplanetary winds}

While our work avoids directly relating our wind prescription to a specific physical mechanism, a proper analysis requires comparing it with the different scenarios that can produce such outflows. As discussed in Sect.~\ref{sec:introduction}, different physical mechanisms may give rise to planetary outflows, most of which are fundamentally driven by accretion processes. In the early stages of planet formation, heating due to pebble accretion can generate a polar "feather" of thermal pressure, which is better observed in \cite{Chrenko2025_CObubble}. While no dedicated study has yet addressed this process in detail, the results of \citet{Chrenko2019,Chrenko2023_AccretingLum,Chrenko2025_CObubble} suggest that such a thermally driven mechanism is a robust mechanism for flow generation and can produce localized outflows extending over a few scale heights. This behavior aligns with the features observed in our $\Gamma_2$ case, although a broader wind structure ($n\sim4$ or even smaller) may offer a better match to the expected geometry.

At later stages, as the planetary core grows and transitions into a phase dominated by gas accretion, magnetic effects become increasingly relevant. The interaction between circumplanetary gas flows and magnetic fields can drive magneto-centrifugal outflows, resulting in more structured and collimated winds. This evolution aligns with the behavior observed in our stronger wind models, $\Gamma_4$ and $\Gamma_8$, which better reflect the large-scale outflows predicted by MHD theory. Supporting this scenario, MHD studies, such as those by \citet{Gressel2013GLOBAL} and \citet{Wafflard-Fernandez2023_planetDiskWindInteraction}, report sustained meridional outflows originating from the conical regions above and below the circumplanetary disk. The emergence of these outflows in studies not specifically designed to produce them shows they are a robust feature of global MHD simulations and likely represent a common outcome of disk-planet interactions. For future modeling efforts, we suggest exploring even higher wind strengths and allowing for non-zero inclination angles (i.e., $\theta_{\mathrm{inc}} \neq 0$) to better capture the stochastic and potentially asymmetric nature of real outflows.

Regardless of the specific launching mechanism, planetary outflows are likely a natural step in the planet formation process. They offer a pathway for angular momentum regulation, envelope mass loss, and modification of circumplanetary dynamics. As shown in our simulations, their impact extends well beyond the planet's immediate vicinity, disrupting polar accretion pathways and potentially altering growth trajectories. Our results show that winds can reduce the mass of the gaseous envelope and limit planetary growth in the context of runaway accretion, and would likely also limit the accretion of small dust particles. However, how the outflows affect solid accretion and their dynamics is yet to be explored.

Because our wind prescription incorporates features common to both magnetically and thermally driven outflows, such as meridional geometry and varying strengths, our results remain broadly applicable across a range of physical scenarios. Future studies that explicitly couple wind launching to detailed physical models will be essential for refining the conclusions presented here. Nevertheless, even in its abstract form, our wind model serves as an effective proxy for the outflow phenomena likely to occur in realistic planet-forming environments.

\begin{acknowledgements}
We acknowledge support from  Agencia Nacional de Investigaci\'on y Desarrollo de Chile (ANID) through QUIMAL fund ASTRO21-0039 and FONDECYT projects 1231205 and 1251456. 
\end{acknowledgements}

%

\bibliographystyle{aa}
\bibliography{bibliography}







    
    



\begin{appendix}




\twocolumn

\section{Wind parameterization} \label{sec:wind_parameterization}
\begin{figure}[!h]
    \centering
    \includegraphics[width=0.98\linewidth]{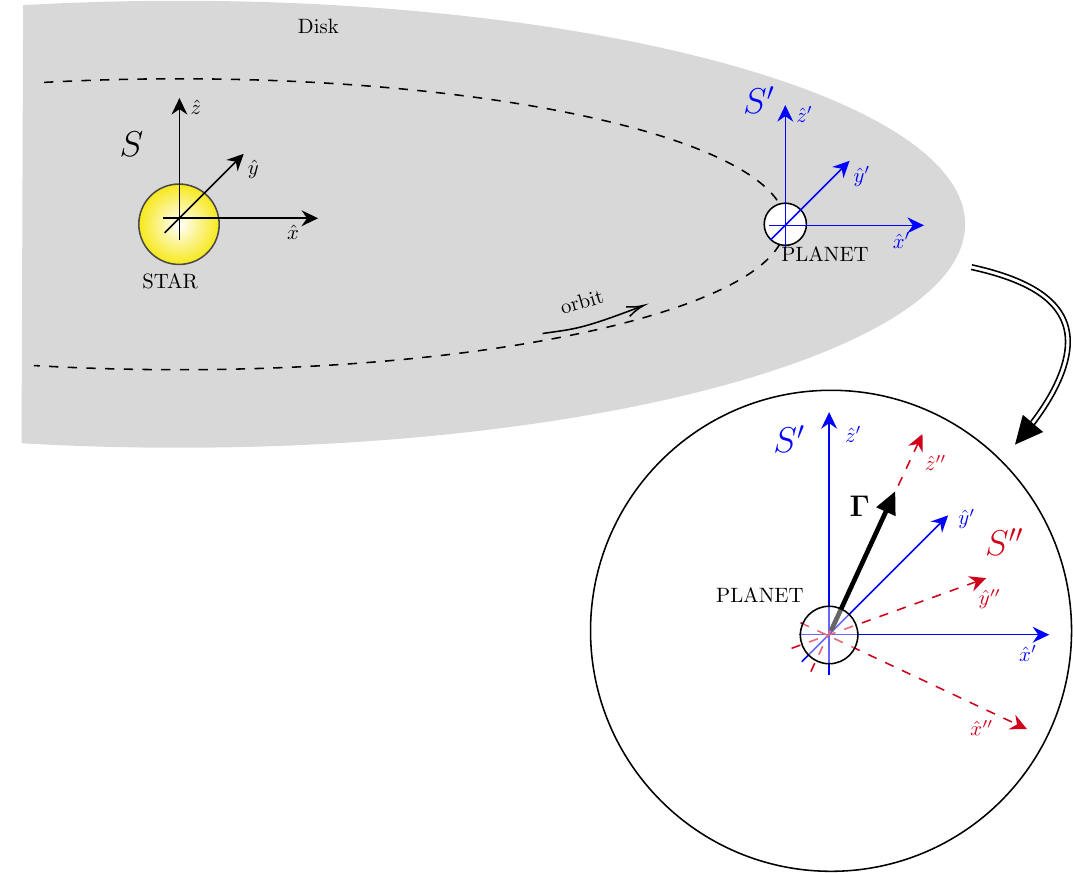}
    \caption{Reference frames used for the wind parameterization. The global frame $\mathbf{S}$ (black) is the inertial frame centered on the star. The co-moving frame $\mathbf{S'}$ (blue) is centered on the planet. The wind is defined in frame $\mathbf{S''}$ (red), which is rotated from $\mathbf{S'}$ so that its $\hat{z}^{\prime\prime}$-axis aligns with the wind's primary direction, represented by the vector $\mathbf{\Gamma}$.}
    \label{fig:parametrization_systems}
\end{figure}

To describe this process, we define the following reference frames:

\begin{itemize}
    \item $\mathbf{S}$: System fixed in the center of the disk. This is the system defined by \texttt{FARGO3D} and used in the simulations.

    \item $\mathbf{S'}$: System fixed at the position of the planet.

    \item $\mathbf{S''}$: System rotated with respect to $\mathbf{S'}$, and where the wind is generated.
\end{itemize}
A sketch of these reference frames is given in Fig.~\ref{fig:parametrization_systems}.

To implement the wind parameterization, we must transform quantities between the global reference frame $\mathbf{S}$ and the local wind-generation frame $\mathbf{S''}$. Assuming an arbitrary wind acceleration of the form
\begin{equation}
    \mathbf{\Gamma} = \Gamma(r^{\prime\prime}, \theta^{\prime\prime}, \varphi^{\prime\prime}) \hat{r}^{\prime\prime},
\end{equation}
defined in $\mathbf{S''}$, we require expressions for the coordinates $(r'', \theta'', \varphi'')$ as functions of the coordinates in the global frame $(r, \theta, \varphi)$ in order to evaluate $\Gamma$. Additionally, since the hydrodynamic solver operates in frame $\mathbf{S}$, the vector $\mathbf{\Gamma}$ (with $\hat{r}^{\prime\prime}$ direction) must be decomposed into components along the basis vectors $(\hat{r}, \hat{\theta}, \hat{\varphi})$. The mathematical details of this transformation between systems are provided in Appendix \ref{sec:wind_parameterization}.

\subsection{Rotation matrices}
Rotation matrices are linear transformations that describe the rotation of an object in Euclidean space. For a rotation by an angle $\alpha$ about the $\hat{z}$-axis in a Cartesian coordinate system, the transformation is given by the matrix $R_z(\alpha)$:
\begin{equation}
    R_z(\alpha) = \begin{pmatrix}
    \cos{\alpha} & -\sin{\alpha} & 0 \\
    \sin{\alpha} & \cos{\alpha} & 0 \\
    0 & 0 & 1 \\
    \end{pmatrix}.
    \label{appendix-eq: R_z}
\end{equation}

This matrix rotates the $xy$-plane by an angle $\alpha$. Similarly, rotations around the $\hat{x}$ and $\hat{y}$ axes are described by $R_x(\alpha)$ and $R_y(\alpha)$, respectively:

\begin{align}
    R_x(\alpha) = \begin{pmatrix}
    1 & 0 & 0 \\
    0 & \cos{\alpha} & -\sin{\alpha} \\
    0 & \sin{\alpha} & \cos{\alpha} \\
    \end{pmatrix},
    \label{appendix-eq: R_x}
    \\
    R_y(\alpha) = \begin{pmatrix}
    \cos{\alpha} & 0 & \sin{\alpha} \\
    0 & 1 & 0 \\
    -\sin{\alpha} & 0 & \cos{\alpha} \\
    \end{pmatrix}.
    \label{appendix-eq: R_y}
\end{align}
Any arbitrary rotation in 3D space can be described by a composition of these fundamental rotation matrices.

\subsection{From stellar to planetary frame ($\mathbf{S} \rightarrow \mathbf{S'}$)}

The transformation from the inertial stellar frame, $\mathbf{S}$, to the co-moving planetary frame, $\mathbf{S'}$, is a simple translation. The origin of $\mathbf{S'}$ orbits the star at a radius $R_0$ with an angular velocity $\omega_0$. The coordinate transformation is:

\begin{equation}
    \begin{aligned}
        x^{\prime} &= x - r_0\cos{\omega_0 t} \\
        y' &= y - r_0 \sin{\omega_0 t} \\
        z' &= z. \\
    \end{aligned}
    \label{eq:WindParametrization-StoS1-coords}
\end{equation}
By construction, the axes of the two frames are parallel:

\begin{equation}
    \begin{aligned}
        \bm{\hat{x}}' &= \bm{\hat{x}} \\
        \bm{\hat{y}}' &= \bm{\hat{y}} \\
        \bm{\hat{z}}' &= \bm{\hat{z}}
    \end{aligned}
    \label{eq:WindParametrization-StoS1-hat}
\end{equation}
which simplifies the transformation between the two frames.

\subsection{From planetary to wind frame ($\mathbf{S'} \rightarrow \mathbf{S''}$)}

The wind source, $\mathbf{\Gamma}$, is defined in the $\mathbf{S''}$ frame, which is rotated with respect to the planetary frame $\mathbf{S'}$. This rotation is constructed by applying a sequence of elementary rotations:

\begin{itemize}
    \item Precession: A rotation by an angle $\varphi_{\text{prec}}$ around the $\bm{\hat{z}'}$-axis, described by the matrix $R_z(\varphi_{\text{prec}})$.

    \item Inclination: A tilt by an angle $\theta_{\text{inc}}$ around the new $\bm{\hat{y}'}$-axis, described by $R_y(\theta_{\text{inc}})$. This orients the wind's primary axis.

    \item Spin: An additional rotation, $R_{\text{spin}}$, can be applied around any axis in the new frame to account for other rotational degrees of freedom. In this work, we do not apply a spin, so $R_{\text{spin}}$ is the identity matrix.
\end{itemize}
Note that all rotation angles defined above could vary with time if needed.

\begin{figure}
    \centering
    \includegraphics[width=0.8\columnwidth]{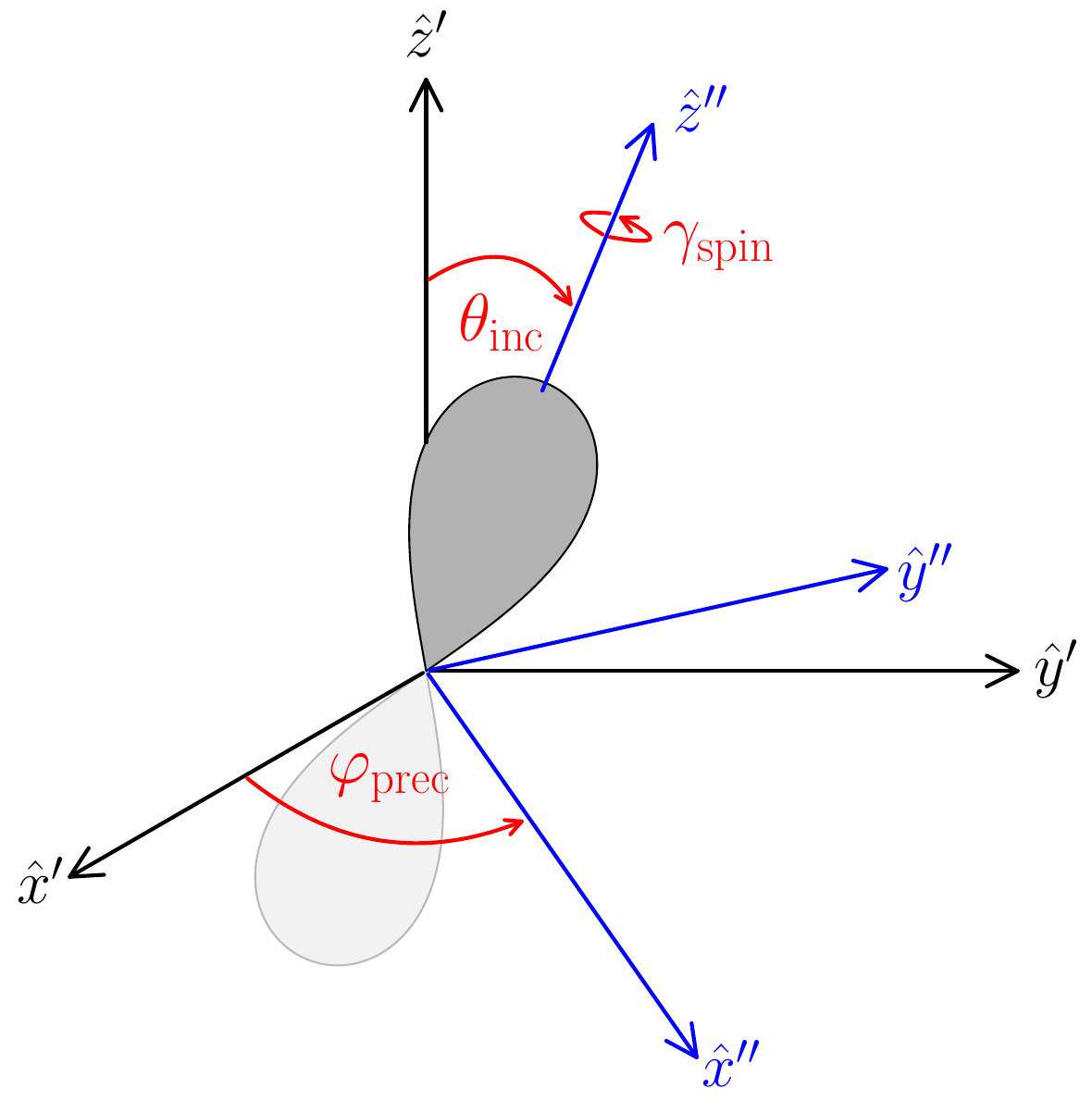}
    \caption{Transformation from System S' to S'' through Precession ($\varphi_{\text{prec}}$), Inclination ($\theta_{\text{inc}}$), and Spin ($\gamma_{\text{spin}}$).}
    \label{fig:S_prima-to-S_prima2}
\end{figure}
The final rotation matrix, $\mathcal{R}$, is the product of these individual rotations, applied sequentially:

\begin{equation} \label{eq:WindParametrization-rotationMatrixS2}
    \mathcal{R} = R_z(\varphi_{\text{prec}}) R_y(\theta_{\text{inc}}) R_{\text{spin}},
\end{equation}
where $R_z(\varphi_{\text{prec}})$ and $R_y(\theta_{\text{inc}})$ are given in Eqs.~\ref{eq:Precession-Rmatrix} and \ref{eq:Inclination-Rmatrix}, respectively. $R_{\text{spin}}$ is not shown as is not applied in this work, but it could be used to incorporate any other desired rotation of $\mathbf{\Gamma}$.

\begin{equation} \label{eq:Precession-Rmatrix}
    R_z(\varphi_{\text{prec}}) = \begin{pmatrix}
        \cos{\varphi_{\text{prec}}} & -\sin{\varphi_{\text{prec}}} & 0 \\
        \sin{\varphi_{\text{prec}}} & \cos{\varphi_{\text{prec}}} & 0 \\
        0 & 0 & 1 \\
    \end{pmatrix},
\end{equation}

\begin{equation} \label{eq:Inclination-Rmatrix}
    R_y(\theta_{\text{inc}}) = \begin{pmatrix}
        \cos{\theta_{\text{inc}}} & 0 & \sin{\theta_{\text{inc}}} \\
        0 & 1 & 0 \\
        -\sin{\theta_{\text{inc}}} & 0 & \cos{\theta_{\text{inc}}} \\
    \end{pmatrix}.
\end{equation}
The full coordinate transformation from the stellar frame $\mathbf{S}$ to the wind frame $\mathbf{S''}$ is then:

\begin{equation}\label{eq:WindParametrization-s_prima2}
    \begin{bmatrix}
        x^{\prime \prime} \\
        y^{\prime \prime} \\
        z^{\prime \prime} \\
    \end{bmatrix} = \mathcal{R} \begin{bmatrix}
        x^{\prime} \\
        y^{\prime} \\
        z^{\prime} \\
    \end{bmatrix} = \mathcal{R} \begin{bmatrix}
        x - r_p\cos{\omega_pt} \\
        y - r_p\sin{\omega_pt} \\
        z \\
    \end{bmatrix}
\end{equation}
and for the basis vectors:

\begin{equation}\label{eq:WindParametrization-s_prima2_tongo}
    \begin{bmatrix}
        \bm{\hat{x}^{\prime \prime}} \\
        \bm{\hat{y}^{\prime \prime}} \\
        \bm{\hat{z}^{\prime \prime}} \\
    \end{bmatrix} = \mathcal{R} \begin{bmatrix}
        \bm{\hat{x}^{\prime}} \\
        \bm{\hat{y}^{\prime}} \\
        \bm{\hat{z}^{\prime}} \\
    \end{bmatrix} = \mathcal{R} \begin{bmatrix}
        \bm{\hat{x}} \\
        \bm{\hat{y}} \\
        \bm{\hat{z}} \\
    \end{bmatrix}
\end{equation}

Since the transformation from $\mathbf{S'}$ to $\mathbf{S''}$ is a pure rotation, the radial distance from the planet remains invariant ($r^{\prime \prime} = r^{\prime}$). The spherical coordinates in the wind frame are:

\begin{subequations}
\begin{align}
        r^{\prime \prime} &= \sqrt{x^{\prime \prime 2} + y^{\prime \prime 2} + z^{\prime \prime 2}} = \sqrt{x^{\prime 2} + y^{\prime 2} + z^{\prime 2}} \\
        \theta^{\prime \prime} &= \arccos{\frac{z^{\prime \prime}}{r^{\prime \prime}}} \\
        \varphi^{\prime \prime} &= \text{atan2}(y^{\prime \prime}, x^{\prime \prime})
\end{align}
\end{subequations}

With these transformations, we can evaluate the wind acceleration $\mathbf{\Gamma}(r^{\prime \prime}, \theta^{\prime \prime}, \varphi^{\prime \prime})$ at any point in the simulation domain.

Finally, we must express the wind's radial unit vector, $\bm{\hat{r}''}$, in terms of the basis vectors of the global spherical coordinate system $(\bm{\hat{r}}, \bm{\hat{\theta}}, \bm{\hat{\varphi}})$. Using Eq.~\eqref{eq:WindParametrization-s_prima2_tongo}, we first write $\bm{\hat{r}''}$ in Cartesian components within the stellar frame:

\begin{equation}\label{eq:WindParametrization-rpima2_cartesian}
\begin{aligned}
    \bm{\hat{r}}^{\prime \prime} &= \sin{\theta^{\prime \prime}}\cos{\varphi^{\prime \prime}}\bm{\hat{x}}^{\prime \prime} + \sin{\theta^{\prime \prime}}\sin{\varphi^{\prime \prime}}\bm{\hat{y}}^{\prime \prime} + \cos{\theta^{\prime \prime}}\bm{\hat{z}}^{\prime \prime} \\
    &= \begin{bmatrix}
        \sin{\theta^{\prime \prime}}\cos{\varphi^{\prime \prime}} \\
        \sin{\theta^{\prime \prime}}\sin{\varphi^{\prime \prime}} \\
        \cos{\theta^{\prime \prime}} \\
    \end{bmatrix}^T
    \begin{bmatrix}
        \bm{\hat{x}^{\prime \prime}} \\
        \bm{\hat{y}^{\prime \prime}} \\
        \bm{\hat{z}^{\prime \prime}} \\
    \end{bmatrix} \\
    &= \begin{bmatrix}
        \sin{\theta^{\prime \prime}}\cos{\varphi^{\prime \prime}} \\
        \sin{\theta^{\prime \prime}}\sin{\varphi^{\prime \prime}} \\
        \cos{\theta^{\prime \prime}} \\
    \end{bmatrix}^T \mathcal{R} \begin{bmatrix}
        \bm{\hat{x}} \\
        \bm{\hat{y}} \\
        \bm{\hat{z}} \\
    \end{bmatrix}.
\end{aligned}
\end{equation}

Next, we transform the Cartesian basis vectors $(\bm{\hat{x}}, \bm{\hat{y}}, \bm{\hat{z}})$ into the spherical basis vectors of the simulation frame using the standard transformation:

\begin{equation}\label{eq: cartesian2spherical}
    \begin{bmatrix}
        \bm{\hat{x}} \\
        \bm{\hat{y}} \\
        \bm{\hat{z}} \\
    \end{bmatrix} = \begin{pmatrix}
        \sin{\theta}\cos{\varphi} & \cos{\theta}\cos{\varphi} & -\sin{\varphi}\\
        \sin{\theta}\sin{\varphi} & \cos{\theta}\sin{\varphi} & \cos{\varphi}\\
        \cos{\theta} & -\sin{\theta} & 0\\
    \end{pmatrix} \begin{bmatrix}
        \bm{\hat{r}} \\
        \bm{\hat{\theta}} \\
        \bm{\hat{\varphi}} \\
    \end{bmatrix}.
\end{equation}

Substituting this into Eq.~\eqref{eq:WindParametrization-rpima2_cartesian}, we obtain the final expression for $\bm{\hat{r}^{\prime \prime}}$ in the simulation's spherical coordinate system:

\begin{align} \label{eq:rTongo-prima2}
    \bm{\hat{r}^{\prime \prime}} =& \begin{bmatrix}
        \sin{\theta^{\prime \prime}}\cos{\varphi^{\prime \prime}} \\
        \sin{\theta^{\prime \prime}}\sin{\varphi^{\prime \prime}} \\
        \cos{\theta^{\prime \prime}}
    \end{bmatrix}^T \notag \\ 
    & \times \mathcal{R} \begin{pmatrix}
        \sin{\theta}\cos{\varphi} & \cos{\theta}\cos{\varphi} & -\sin{\varphi}\\
        \sin{\theta}\sin{\varphi} & \cos{\theta}\sin{\varphi} & \cos{\varphi}\\
        \cos{\theta} & -\sin{\theta} & 0\\   
    \end{pmatrix}
    \begin{bmatrix}
        \bm{\hat{r}} \\
        \bm{\hat{\theta}} \\
        \bm{\hat{\varphi}} \\
    \end{bmatrix}.
\end{align}

This equation allows us to compute the components of the wind acceleration vector $\mathbf{\Gamma} = \Gamma(r^{\prime \prime}, \theta^{\prime \prime}, \varphi^{\prime \prime})\bm{\hat{r}^{\prime \prime}}$ in the global spherical frame $(\bm{\hat{r}}, \bm{\hat{\theta}}, \bm{\hat{\varphi}})$ at every point in the grid.

\section{Choosing the strength of the wind} \label{section:Strength parameter}

\begin{figure*}[!h]
    \centering
    \includegraphics[width=0.75\textwidth]{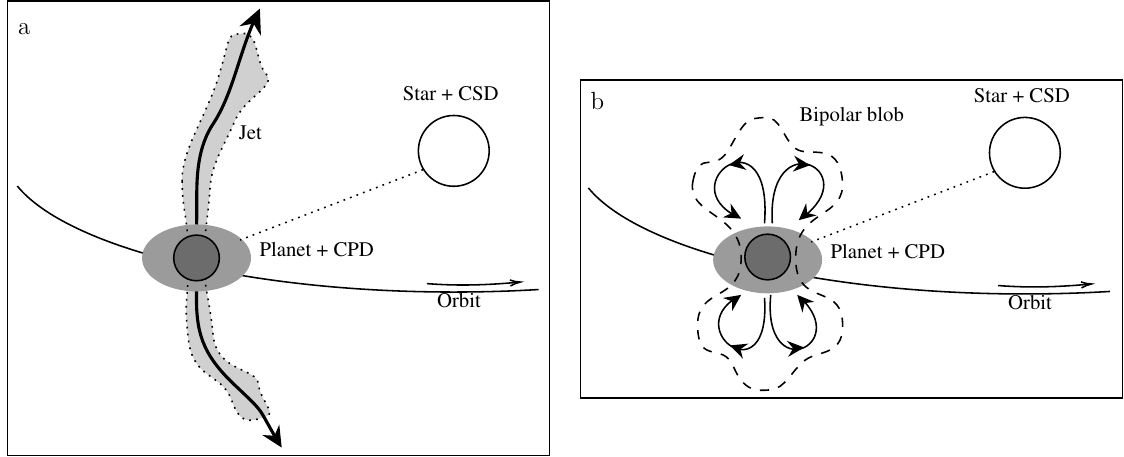}
    \caption{Illustration of the two planet-driven outflow regimes (based on \cite{Fendt2003}, Figure 1), showing a planet embedded in a circumstellar disk (CSD). (a) Jet: Outflows exceeding the escape velocity ($v_w > v_e$) escape the circumplanetary disk (CPD). (b) Bipolar blob: Sub-escape-velocity outflows ($v_w < v_e$) remain gravitationally bound to the planet as it orbits. The bipolar blobs are reshaped (with respect to \cite{Fendt2003}) to better match the observed structure of weak outflows in numerical models \citep[e.g.,][]{Chrenko2019,Chrenko2025_CObubble}}
    \label{fig:IllustrationWindsVsGp}
\end{figure*}

In this work, we investigate the role of planet-driven winds across different spatial scales. \cite{Fendt2003} analytically studied the conditions for MHD-driven planetary outflows and their dynamical effects. While the driving mechanism in our work differs, their framework provides a valuable foundation for interpreting outflow regimes. In particular, \cite{Fendt2003} identified two distinct morphologies, which are illustrated in Fig.~\ref{fig:IllustrationWindsVsGp}:

\begin{itemize}
    \item Jets: High-velocity outflows where the wind speed exceeds the planet's escape velocity ($v_w > v_e$), enabling material to escape the circumplanetary disk (CPD).
    \item Bipolar blobs: Sub-critical outflows ($v_w < v_e$) where material remains gravitationally bound to the planet, forming transient co-orbiting structures.
\end{itemize}

Although \cite{Fendt2003} concluded that jets naturally escape from the disk (neglecting hydrodynamic drag) based on the finding that the planetary escape velocity exceeds the stellar one in typical scenarios, we will show later that this result does not directly extend to our parameterized wind model. The key differences lie in (1) the nature of the wind-driving mechanism, and (2) the spatial extent of our setup, which applies winds out to $r_{\mathrm{H}}/2$, significantly beyond the small spatial scales (comparable to the planetary radius) considered in \cite{Fendt2003}.

There are two primary conditions that are required to produce jet-like outflows:

\begin{itemize}
    \item Local wind dominance: the wind-induced acceleration must locally overcome planetary gravity, i.e., $\Gamma(r')/G_p(r') > 1$ at radius $r'$.
    
    \item Escape criterion: the parcel must gain sufficient velocity to escape the planetary potential, i.e., $v_w(r', r_{\mathrm{H}}/2) > v_e(r_{\mathrm{H}}/2)$.
\end{itemize}

Neglecting the angular component, $\theta''$, from Eq.~\eqref{eq:GammaDef} for this analysis, it can be shown that the first condition reads:

\begin{equation}
    A_{\gamma} \left( \frac{M_{\star}}{9M_p} \right)^{1/3} \left( \frac{r'}{r_{\mathrm{H}}} \right)^{2} e^{-r'^2/r_s^2} > 1.
    \label{eq:local_wind_dominance}
\end{equation}
For the second condition, the escape velocity of the planet is given by

\begin{equation}
    v_e(r') = \sqrt{\frac{2GM_p}{r'}}.
\end{equation}
Assuming the parcel starts at radius $r'$ and is accelerated outward, it reaches radius $r_2 = r_{\mathrm{H}}/2$ with velocity $v_2=v_w(r', r_2)$, determined by equating the work done by the net acceleration to its kinetic energy,

\begin{equation}
    \frac{1}{2}v_w^2(r', r_{\mathrm{H}}/2) = \int_{r'}^{r_{\mathrm{H}}/2} 
        \left[ A_{\gamma} \frac{GM_\star}{R_0^2} e^{-x^2/r_s^2} 
        - \frac{GM_p}{x^2} \right] dx
    \label{eq:energy_balance}
\end{equation}
which, after integration, gives the velocity of the parcel as
\begin{equation}
    v_w(r', r_{\mathrm{H}}/2) = \sqrt{ A_{\gamma} \frac{GM_\star r_s \sqrt{\pi}}{R_0^2} b_{\mathrm{erf}}(r') 
        + \frac{4GM_p}{r_{\mathrm{H}}} - \frac{2GM_p}{r'} }
    \label{eq:velocity_raw}
\end{equation}
where we define

\begin{equation*}
        b_{\mathrm{erf}}(r') \equiv \mathrm{erf}\left(\frac{r_{\mathrm{H}}}{2r_s}\right) - \mathrm{erf}\left(\frac{r'}{r_s}\right)
\end{equation*}
to encapsulate the error function (erf) dependence.

Thus, the final expression that represents the second condition for jet-like behavior reads
\begin{equation}
    \sqrt{ A_{\gamma} C(r') + 1 - \frac{r_{\mathrm{H}}}{2r'} } > 1,
    \label{eq:escape_condition}
\end{equation}
where $C(r')$ is a dimensionless coefficient that groups physical parameters as:

\begin{equation}
\begin{aligned}
    C(r') & = \left(\frac{M_\star}{9M_p}\right)^{1/3} \frac{r_s \sqrt{\pi}}{4 r_{\mathrm{H}}} b_{\text{erf}}(r')\\ 
    &= \left(\frac{M_\star}{9M_p}\right)^{1/3} \frac{\sqrt{\pi}}{8} b_{\text{erf}}(r') \quad \text{($r_s = r_{\mathrm{H}}/2$)}.
\end{aligned}
\end{equation}

\begin{figure}[!ht]
    \centering
    \includegraphics[width=0.95\columnwidth]{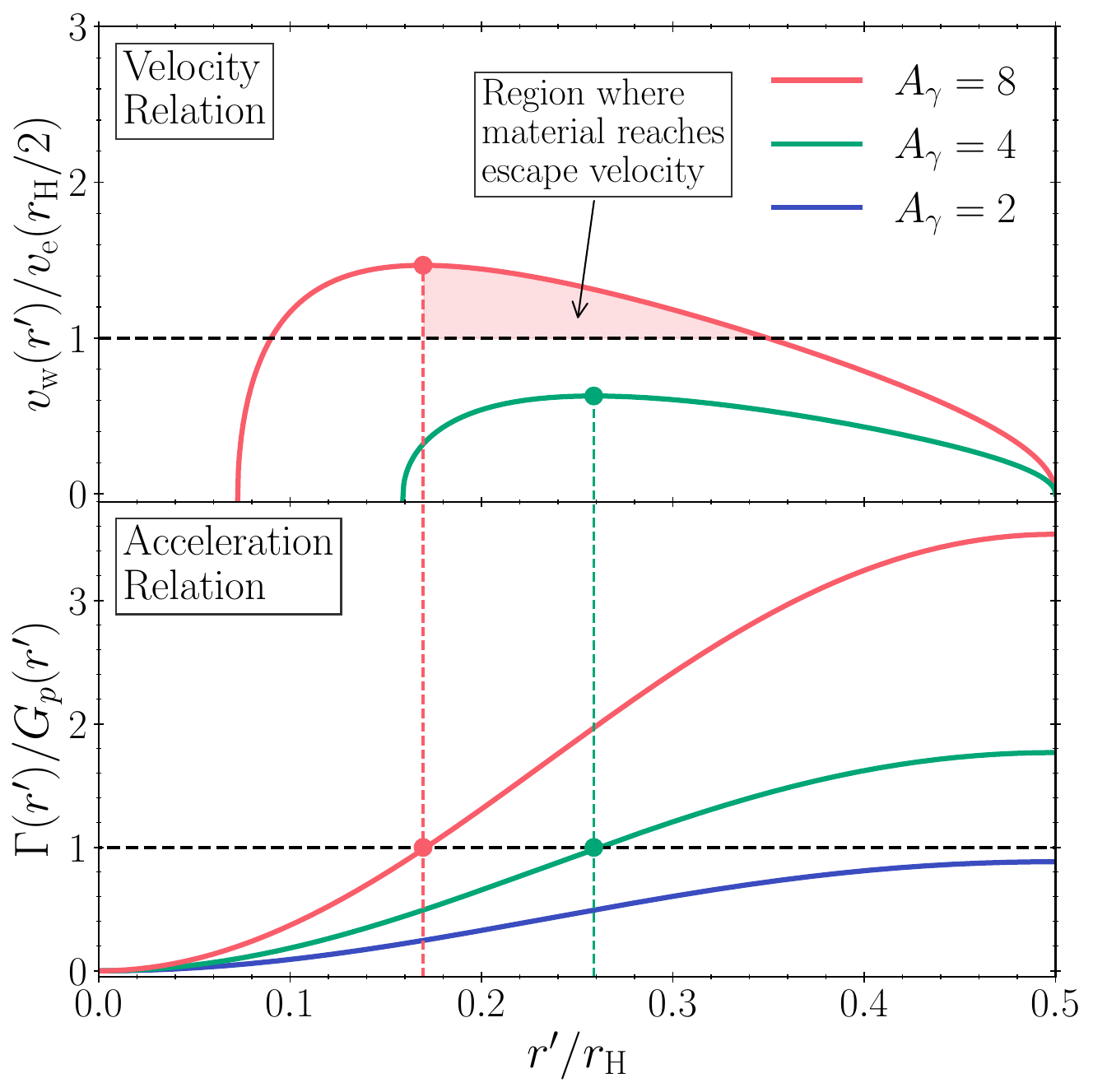}
    \caption{
        Numerical analysis of the conditions for jet-like outflows for $A_{\gamma}=2,4,8$ (blue, green, and red). The bottom panel shows condition~1 (Eq.~\eqref{eq:local_wind_dominance}), which is satisfied beyond the radii marked by the vertical dashed lines for $A_{\gamma}=4$ and~8. The top panel shows condition~2 (Eq.~\eqref{eq:escape_condition}); the horizontal dashed line indicates the escape velocity threshold $v_w(r',r_{\mathrm{H}}/2)/v_e(r_{\mathrm{H}}/2)=1$. The shaded red region highlights the radial range where both conditions are simultaneously satisfied for $A_{\gamma}=8$, identifying the region from which material is ejected from the planet's vicinity.}
    \label{fig:wind_escape_analysis}
\end{figure}

Figure \ref{fig:wind_escape_analysis} illustrates the interplay between these two conditions. For instance, when $A_{\gamma} = 4$, Condition 1 is satisfied near $r' \approx 0.26r_{\mathrm{H}}$ (bottom panel), yet fails Condition 2 (top panel). In contrast, when $A_{\gamma} = 8$, both conditions are satisfied for $0.17r_{\mathrm{H}} \lesssim r' \lesssim  0.35r_{\mathrm{H}}$ (shaded region), enabling sustained jet-like behavior. Because $A_{\gamma} = 2, 4,$ and $8$ cover different outflow regimes, we have chosen these values for our simulations.

We also compare the escape velocity for the planet, $v_{e,p}(r_{\mathrm{H}}/2)$, with that of the star, $v_{e,\star}(R_0)$:

\begin{equation}
 \frac{v_{e,p}(r_{\mathrm{H}}/2)}{v_{e,\star}(R_0)} = (24)^{1/6} \left(\frac{M_p}{M_{\star}} \right)^{1/3} \approx 1.7 \left(\frac{M_p}{M_{\star}} \right)^{1/3}.
\end{equation}
Using a Jupiter-like planet-star relation ($M_p/M_{\star} \sim 10^{-3}$), we get an escape velocity ratio of about 0.17, which leads to the conclusion that exceeding the planet's escape velocity does not mean that material is escaping from the disk. Under the same conditions, for the results of \cite{Fendt2003} to hold ($v_{e,p} > v_{e,\star}$), we would have had to set $r_s < r_{\mathrm{H}}/24$, which is extremely computationally expensive for a grid-based code.


\section{Accretion pathways} \label{section:appendix_accretionPathways}
\begin{figure}
    \centering
    \includegraphics[width=\columnwidth]{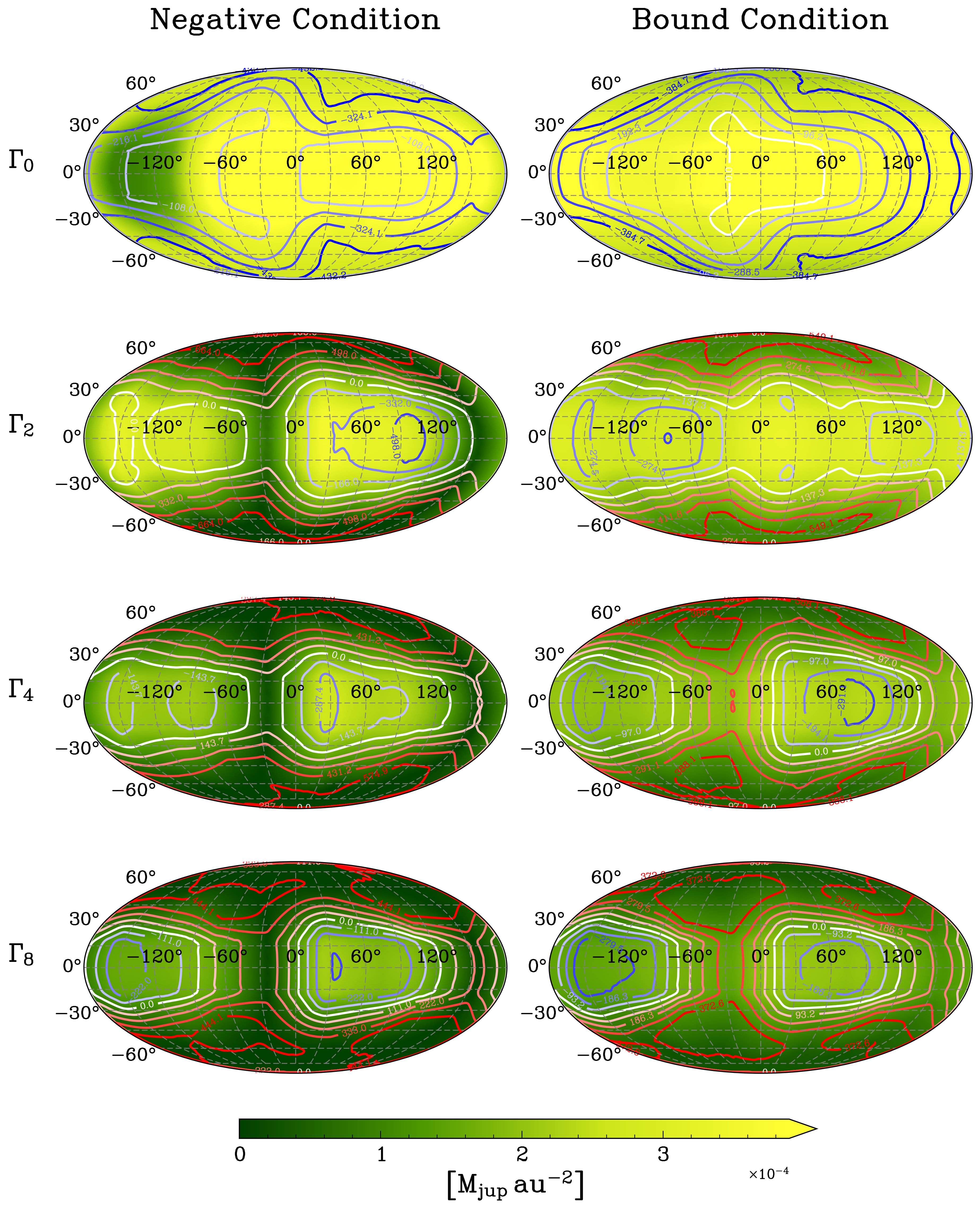}
    \caption{Radial slice of the accreted mass surface density at 50 orbits for $\kappa=5$. Rows correspond to the models $\Gamma_0$, $\Gamma_2$, $\Gamma_4$, and $\Gamma_8$ from top to bottom. The left column shows simulations employing the negative radial velocity condition, while the right column shows simulations using the bound material condition. The maps are extracted on a spherical shell at $R=0.15\,r_{\mathrm{H}}$. The color scale indicates the accreted mass per unit area in units of $M_{\mathrm{jup}}\,\mathrm{au}^{-2}$. Overlaid contours show the mass flux, with signs defined according to Eq.~\eqref{eq:massFluxDef}.}
    \label{fig:projectionMap_macc_kappa05}
\end{figure}

\begin{figure}
    \centering
    \includegraphics[width=\columnwidth]{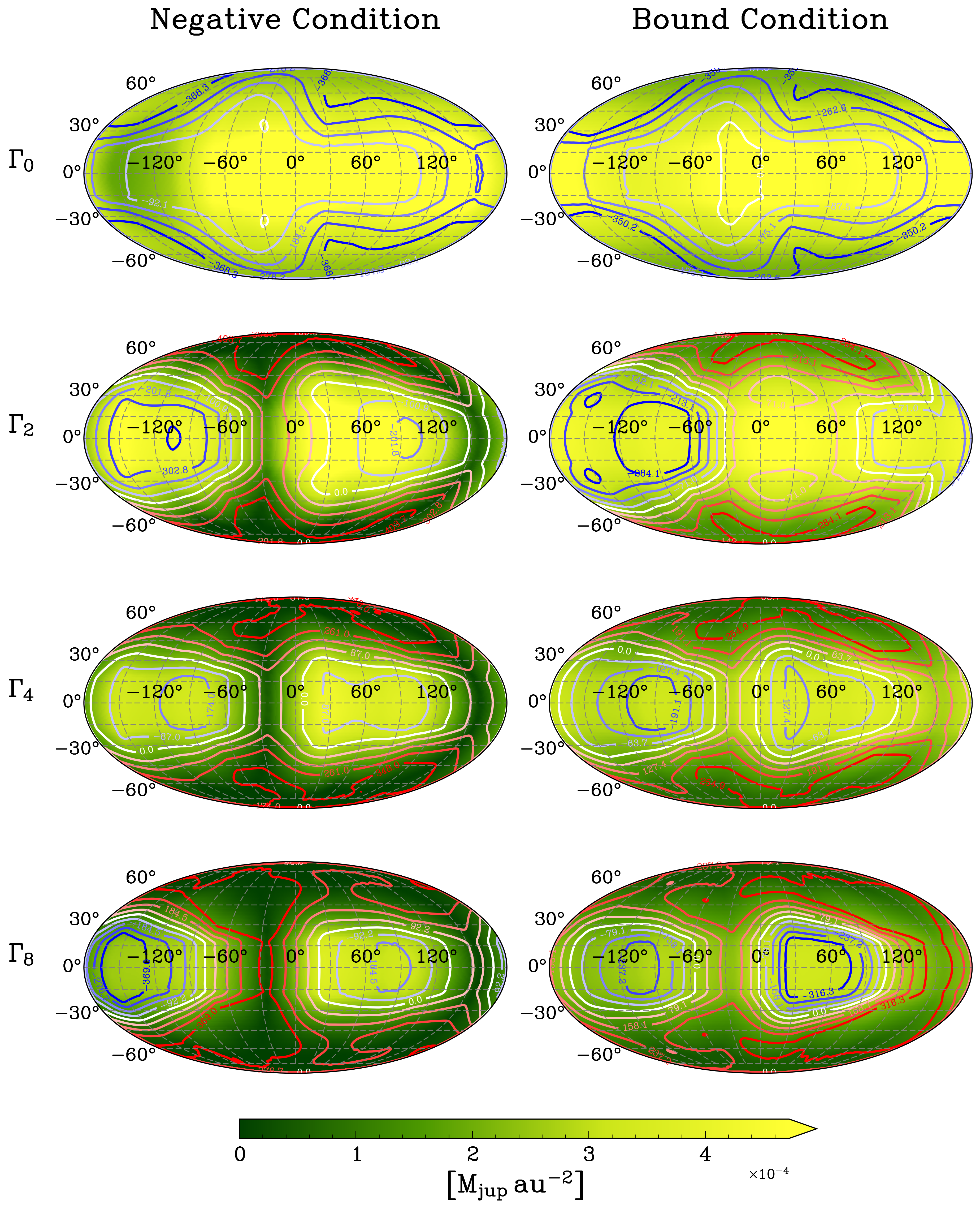}
    \caption{Same as Figure~\ref{fig:projectionMap_macc_kappa05}, but for $\kappa=10$.}
    \label{fig:projectionMap_macc_kappa10}
\end{figure}

\begin{figure}
    \centering
    \includegraphics[width=\columnwidth]{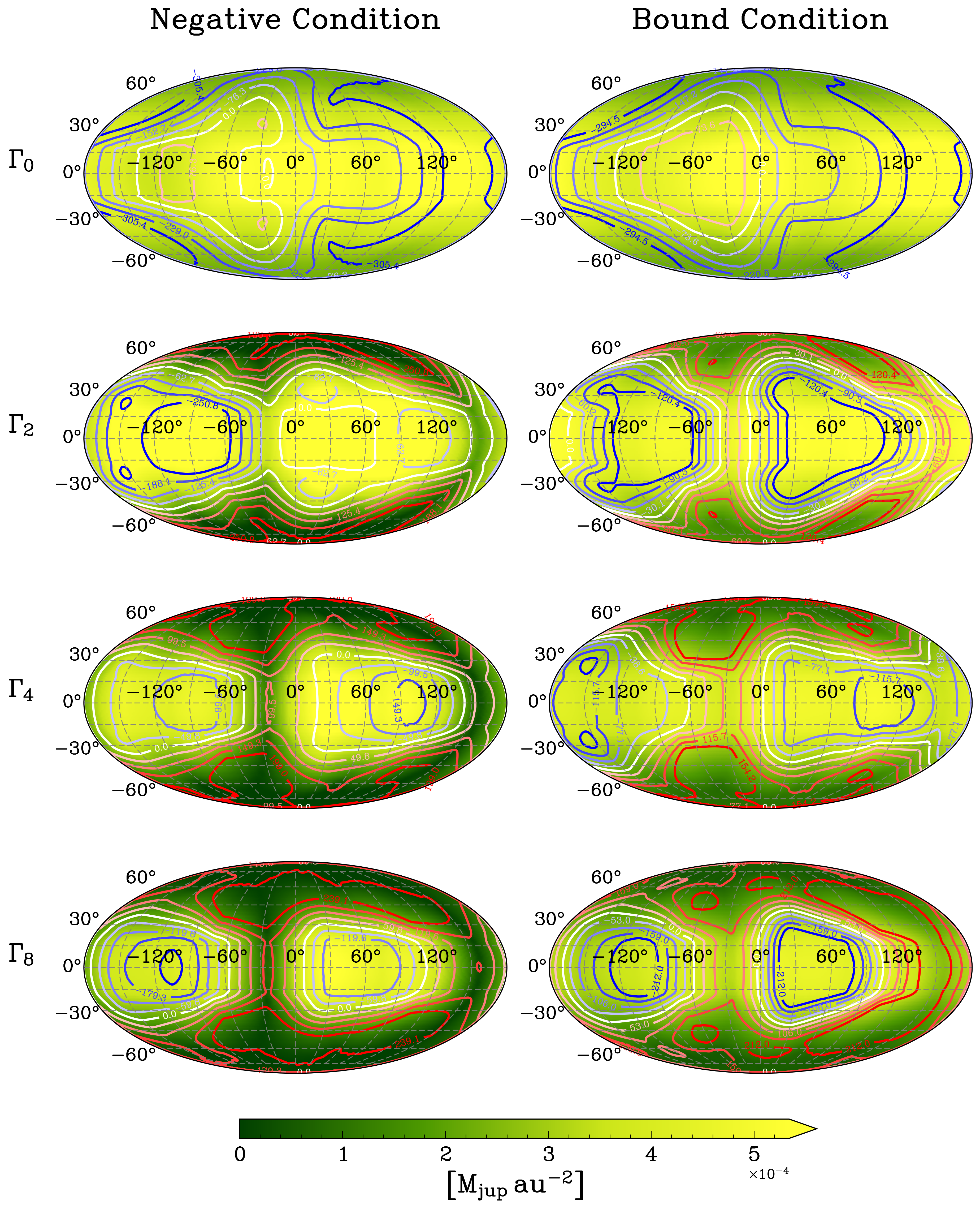}
    \caption{Same as Figure~\ref{fig:projectionMap_macc_kappa05}, but for $\kappa=20$.}
    \label{fig:projectionMap_macc_kappa20}
\end{figure}

To perform a detailed analysis of the accretion pathways in our simulations, we generated accretion maps and flux projections in the close-in regions of the planet for all parameter configurations. The results are interpolated to compensate for the limited resolution in this region. We chose a projection radius of $R=0.15\,r_{\mathrm{H}}$. Figures~\ref{fig:projectionMap_macc_kappa05}, \ref{fig:projectionMap_macc_kappa10}, and \ref{fig:projectionMap_macc_kappa20} show the radial projections of the accretion rate at 50 orbits for different values of $\kappa$ across all configurations.

We observe clear differences in the magnitude and spatial distribution of accreted material when comparing simulations with different accretion efficiencies ($\kappa$). These variations correlate with the total accreted mass trends shown in Fig.~\ref{fig:accretion_analysis}. In the baseline models ($\Gamma_0$), accretion occurs from various directions. Although the limited resolution prevents the identification of a single predominant pathway, accretion appears to proceed comparably from both the midplane and polar regions, resulting in a more isotropic pattern. For lower efficiencies ($\kappa=5$ and $10$), the $\Gamma_0$ simulations exhibit projection features in the same regions, which are more pronounced at lower $\kappa$. This is attributed to the presence of horseshoe flows, which naturally drive material inward and outward; this effect diminishes as accretion becomes more efficient.

In wind-active simulations, the wind significantly alters the accretion pathways. Regions primarily impacted by the outflow act as a barrier to accretion. Specifically, polar regions show minimal accretion compared to the midplane, an effect that becomes more pronounced with stronger winds. The wind also constricts the region near the midplane where accretion can occur, particularly under the negative radial velocity condition. Consequently, as wind strength increases, the accretion window narrows, reinforcing the shift toward midplane-dominated accretion.

The divergence between the two accretion criteria (bound material vs. negative radial velocity) becomes more noticeable in the presence of a wind. The negative radial velocity condition is more restrictive regarding polar accretion, especially for stronger winds. In general, these two conditions complement each other; physically, we expect a combination where gravitationally bound material eventually infalls, while strong supersonic inflows also contribute. However, under both conditions, we observe that accretion in wind-active models is primarily driven by midplane flows, consistent with the latitudinal mass flux analysis presented in Fig.~\ref{fig:avgf_by_lat}.

\FloatBarrier 
\clearpage

\end{appendix}
\end{document}